\documentclass[12pt,a4paper,oneside]{article}
\usepackage[affil-it]{authblk}
\usepackage[centertags]{amsmath}
\usepackage{amssymb,mathtools,mathrsfs}
\usepackage{amsthm}
\usepackage{enumerate}
\usepackage{stmaryrd}
\usepackage{dsfont}
\usepackage{url}
\usepackage{graphicx}
\usepackage{color}
\usepackage{epsfig}
\usepackage{subcaption}
\usepackage{tikz}
\usepackage{hyperref}
\usepackage[a4paper,tmargin=3cm,bmargin=3cm,rmargin=2.2cm,lmargin=2.2cm]{geometry}

\definecolor{dg}{rgb}{0,0.5,0}		
\definecolor{orange}{rgb}{1.0, 0.5, 0.0}

\usepackage{hyphenat}
\usepackage[english]{babel}
\usepackage[T1]{fontenc}

\theoremstyle{theorem}
\newtheorem{theorem}{Theorem}
\newtheorem{corollary}[theorem]{Corollary}
\newtheorem{lemma}[theorem]{Lemma}
\newtheorem{proposition}[theorem]{Proposition}

\theoremstyle{definition}

\newtheorem{remark}[theorem]{Remark}

\newcommand{\sM}{{\mathbb M}}

\newcommand{\sR}{{\mathbb R}}
\newcommand{\sC}{{\mathbb C}}

\newcommand{\PCn}{\mathbb{P}(\mathbb{C}^n)}

\newcommand{\B}{\mathcal{B}}
\newcommand{\C}{\mathcal{C}}

\newcommand{\K}{\mathcal{K}}

\newcommand{\cO}{\mathcal{O}}
\newcommand{\cS}{\mathcal{S}}
\newcommand{\cX}{\mathcal{X}}

\newcommand{\bjed}{\mathbf{1}}
\newcommand{\bx}{\textup{\textbf{x}}}
\newcommand{\by}{\textup{\textbf{y}}}
\newcommand{\bz}{\textup{\textbf{z}}}

\newcommand{\bDelta}{\boldsymbol\Delta}

\newcommand{\bOmega}{\boldsymbol\Omega}
\newcommand{\bba}{\textup{\textbf{a}}}

\newcommand{\be}{\mathbf{e}}
\newcommand{\bbf}{\mathbf{f}}
\newcommand{\bg}{\mathbf{g}}
\newcommand{\bu}{\mathbf{u}}
\newcommand{\bv}{\mathbf{v}}
\newcommand{\bw}{\mathbf{w}}
\newcommand{\bA}{\mathbf{A}}
\newcommand{\bB}{\mathbf{B}}
\newcommand{\bD}{\mathbf{D}}
\newcommand{\bE}{\textup{\textbf{E}}}
\newcommand{\bG}{\mathbf{G}}
\newcommand{\bH}{\mathbf{H}}
\newcommand{\bI}{\mathbf{I}}
\newcommand{\bM}{\mathbf{M}}
\newcommand{\bP}{\mathbf{P}}
\newcommand{\bQ}{\mathbf{Q}}

\newcommand{\bU}{\mathbf{U}}
\newcommand{\bV}{\mathbf{V}}
\newcommand{\bW}{\mathbf{W}}

\newcommand{\ii}{\textup{i}}
\newcommand{\ee}{\textup{e}}
\newcommand{\dd}{\textup{d}}

\DeclareMathOperator{\diag}{diag}
\DeclareMathOperator{\dist}{dist}

\DeclareMathOperator{\Tr}{Tr}
\DeclareMathOperator{\sgn}{sgn}
\DeclareMathOperator{\cof}{cof}
\DeclareMathOperator{\RE}{Re}
\DeclareMathOperator{\IM}{Im}
\DeclareMathOperator{\sumka}{{\textstyle\sum}}

\newcommand{\sspan}{\textnormal{span}}

\newcommand{\x}{\times}
\newcommand{\w}{\wedge}
\newcommand{\vc}{\vcentcolon =}
\newcommand{\cv}{= \vcentcolon}

\tikzset {_4m3ns55d7/.code = {\pgfsetadditionalshadetransform{ \pgftransformshift{\pgfpoint{89.1 bp } { -108.9 bp }  }  \pgftransformscale{1.32 }  }}}
\pgfdeclareradialshading{_g1pc9w3t7}{\pgfpoint{-72bp}{88bp}}{rgb(0bp)=(1,1,1);
rgb(0bp)=(1,1,1);
rgb(25bp)=(0,0,0);
rgb(400bp)=(0,0,0)}

\tikzset {_jgeqz7du0/.code = {\pgfsetadditionalshadetransform{ \pgftransformshift{\pgfpoint{89.1 bp } { -108.9 bp }  }  \pgftransformscale{1.32 }  }}}
\pgfdeclareradialshading{_qbtkb5o37}{\pgfpoint{-72bp}{88bp}}{rgb(0bp)=(1,1,1);
rgb(0bp)=(1,1,1);
rgb(25bp)=(0,0,0);
rgb(400bp)=(0,0,0)}

\tikzset {_nt16o215o/.code = {\pgfsetadditionalshadetransform{ \pgftransformshift{\pgfpoint{89.1 bp } { -108.9 bp }  }  \pgftransformscale{1.32 }  }}}
\pgfdeclareradialshading{_opmlqqfi0}{\pgfpoint{-72bp}{88bp}}{rgb(0bp)=(1,1,1);
rgb(0bp)=(1,1,1);
rgb(25bp)=(0,0,0);
rgb(400bp)=(0,0,0)}

\tikzset {_yumybsto8/.code = {\pgfsetadditionalshadetransform{ \pgftransformshift{\pgfpoint{89.1 bp } { -108.9 bp }  }  \pgftransformscale{1.32 }  }}}
\pgfdeclareradialshading{_u01x1z4a0}{\pgfpoint{-72bp}{88bp}}{rgb(0bp)=(1,1,1);
rgb(0bp)=(1,1,1);
rgb(25bp)=(0,0,0);
rgb(400bp)=(0,0,0)}

\tikzset {_9uy8x5mc0/.code = {\pgfsetadditionalshadetransform{ \pgftransformshift{\pgfpoint{89.1 bp } { -108.9 bp }  }  \pgftransformscale{1.32 }  }}}
\pgfdeclareradialshading{_itvxi5kv9}{\pgfpoint{-72bp}{88bp}}{rgb(0bp)=(1,1,1);
rgb(0bp)=(1,1,1);
rgb(25bp)=(0,0,0);
rgb(400bp)=(0,0,0)}

\tikzset {_yw3sbxir5/.code = {\pgfsetadditionalshadetransform{ \pgftransformshift{\pgfpoint{89.1 bp } { -108.9 bp }  }  \pgftransformscale{1.32 }  }}}
\pgfdeclareradialshading{_3nmh2rxaf}{\pgfpoint{-72bp}{88bp}}{rgb(0bp)=(1,1,1);
rgb(0bp)=(1,1,1);
rgb(25bp)=(0,0,0);
rgb(400bp)=(0,0,0)}

\tikzset {_03k4z8nuo/.code = {\pgfsetadditionalshadetransform{ \pgftransformshift{\pgfpoint{89.1 bp } { -108.9 bp }  }  \pgftransformscale{1.32 }  }}}
\pgfdeclareradialshading{_fh5lwj2sk}{\pgfpoint{-72bp}{88bp}}{rgb(0bp)=(1,1,1);
rgb(0bp)=(1,1,1);
rgb(25bp)=(0,0,0);
rgb(400bp)=(0,0,0)}

\tikzset {_8fgmug454/.code = {\pgfsetadditionalshadetransform{ \pgftransformshift{\pgfpoint{89.1 bp } { -108.9 bp }  }  \pgftransformscale{1.32 }  }}}
\pgfdeclareradialshading{_fpftkbl0x}{\pgfpoint{-72bp}{88bp}}{rgb(0bp)=(1,1,1);
rgb(0bp)=(1,1,1);
rgb(25bp)=(0,0,0);
rgb(400bp)=(0,0,0)}

\tikzset {_z0xc2egj0/.code = {\pgfsetadditionalshadetransform{ \pgftransformshift{\pgfpoint{89.1 bp } { -108.9 bp }  }  \pgftransformscale{1.32 }  }}}
\pgfdeclareradialshading{_4mn34k8wq}{\pgfpoint{-72bp}{88bp}}{rgb(0bp)=(1,1,1);
rgb(0bp)=(1,1,1);
rgb(25bp)=(0,0,0);
rgb(400bp)=(0,0,0)}

\tikzset {_jv66gayjb/.code = {\pgfsetadditionalshadetransform{ \pgftransformshift{\pgfpoint{89.1 bp } { -108.9 bp }  }  \pgftransformscale{1.32 }  }}}
\pgfdeclareradialshading{_20zjeydf3}{\pgfpoint{-72bp}{88bp}}{rgb(0bp)=(1,1,1);
rgb(0bp)=(1,1,1);
rgb(25bp)=(0,0,0);
rgb(400bp)=(0,0,0)}

\tikzset {_1igd3srsx/.code = {\pgfsetadditionalshadetransform{ \pgftransformshift{\pgfpoint{89.1 bp } { -108.9 bp }  }  \pgftransformscale{1.32 }  }}}
\pgfdeclareradialshading{_kgttk51lb}{\pgfpoint{-72bp}{88bp}}{rgb(0bp)=(1,1,1);
rgb(0bp)=(1,1,1);
rgb(25bp)=(0,0,0);
rgb(400bp)=(0,0,0)}

\tikzset {_nsi30h14y/.code = {\pgfsetadditionalshadetransform{ \pgftransformshift{\pgfpoint{89.1 bp } { -108.9 bp }  }  \pgftransformscale{1.32 }  }}}
\pgfdeclareradialshading{_dym6cyghs}{\pgfpoint{-72bp}{88bp}}{rgb(0bp)=(1,1,1);
rgb(0bp)=(1,1,1);
rgb(25bp)=(0,0,0);
rgb(400bp)=(0,0,0)}
\tikzset{every picture/.style={line width=0.75pt}} 

\begin{document}


\title{Distances between pure quantum states induced~by~a~distance matrix}
\author{Tomasz Miller${}^{1}$\thanks{tomasz.miller@uj.edu.pl}, Rafa{\l} Bistro\'{n}${}^{2,3}$}

\affil{\small ${}^1$ Copernicus Center for Interdisciplinary Studies, Jagiellonian University,
\\Szczepa\'nska 1/5, 31-011 Krak\'ow, Poland \\ $^{2}$ Institute of Theoretical Physics, Jagiellonian University, \\ ul. {\L}ojasiewicza 11, 30--348 Krak\'ow, Poland \\ ${}^{3}$ Doctoral School of Exact and Natural Sciences, Jagiellonian University, \\ ul. Łojasiewicza 11, 30-348 Kraków, Poland}
\date{\today}


\maketitle

\begin{abstract}
Given an $n \times n$ distance matrix, we construct an infinite family of distances $d_p$ (where $p \geq 2$) on the complex projective space $\mathbb{P}(\mathbb{C}^n)$ modelling the space of pure states of an $n$-level quantum system. The construction can be seen as providing a natural way to isometrically embed any given finite metric space into the space of pure quantum states `spanned' upon it. In order to show that the maps $d_p$ are indeed distance functions---in particular, that they satisfy the triangle inequality---we employ methods of analysis, multilinear algebra and convex geometry, obtaining a nontrivial auxiliary convexity result in the process. In addition, a way of extending the distances $d_p$ to mixed states is proposed for a broad class of distance matrices.
\end{abstract}

MSC classes: 54E35, 15A42, 15A69, 52A40, 26B25


\section{Introduction and main result}

The space of pure states of an $n$-level quantum system, identified with the complex projective space\footnote{Often traditionally denoted $\sC\mathbb{P}^N$ where $N=n-1$.} $\PCn$, is a fundamental object of study in quantum mechanics and quantum information theory \cite{BZ17}. Its geometry, particularly the notion of distance between states, has invaluable importance in quantum information processing \cite{Keyl, NielsenChuang}, and is critical for tasks such as quantum metrology and sensing \cite{QSense, QMetro}, or quantum machine learning \cite{QML, KdPMLL22}.

Recall that $\PCn$ can be regarded as the quotient of the unit sphere $\{\bx \in \sC^n \, | \, \|\bx\| = 1\}$ under the action of $\textup{U}(1)$, meaning that one can think of a pure state as a norm-one vector $\bx$ defined up to a phase factor $\ee^{\ii\phi}$, as is customary in physics. 
In quantum information theory, however, pure states are usually modelled as rank-one orthogonal projections, with $\bx \in \PCn$ represented by the matrix\footnote{Throughout the paper, we shall identify vectors with column matrices.} $\bx\bx^\dag$. More generally, quantum states are represented by positive-semidefinite matrices of unit trace, called \emph{density matrices}, with those that are not rank-one called \emph{mixed states}. The set $\Omega_n$ of $n \times n$ density matrices is convex, and its extreme points are precisely the pure states.

The topological space $\PCn$ can be endowed with a metric space structure in many ways. A standard choice of the distance function is the one induced by the Hilbert--Schmidt (or Frobenius) norm on the space of $n \times n$ matrices, namely
\begin{align}\label{eq:HSdist}
d_{\text{HS}}(\bx,\by) \coloneqq \tfrac{1}{\sqrt{2}}\|\bx\bx^\dag - \by\by^\dag\|_\textup{F} = \sqrt{1 - |\bx^\dag \by|^2}
\end{align}
for any $\bx,\by \in \PCn$. While geometrically natural and widely used (along with the Fubini--Study distance given by $d_{\text{FS}}(\bx,\by) = \arccos |\bx^\dag \by| = \arcsin d_{\text{HS}}(\bx,\by)$, cf. \cite{BZ17}), this distance function is not always the most physically motivated measure of proximity. In particular, $d_{\text{HS}}$ assigns the same unit distance to all pairs of orthogonal states. To impose more sophisticated geometry on $\PCn$ and entire $\Omega_n$, several authors introduced distances based on the {\sl quantum transport problem} (see \cite{Beatty,Invitation} for an overview of the current state of the art), which generalized the classical transport problem by Kantorovich \cite{Kan48}, offering various quantum counterparts of the Wasserstein distances \cite{BeattyFranca,CGP20,CarlenMaas,PMTL21,PT21,Duv22,GP18,TothPitrik23,TothPitrik26,ZYYY22}. In this work, we analyze distances directly connected to the quantum optimal transport problem, see \cite{BEZ22,CEFZ21,FECZ21,ZS01}, for pure quantum states.

More concretely, in \cite{FECZ21} the following definition of the quantum $p$-Wasserstein semidistance was put forward. For any two density matrices $\rho_1,\rho_2 \in \Omega_n$ consider the set of their \emph{couplings}, defined as states $\rho_{12} \in \Omega_{n^2}$ whose marginals coincide with $\rho_1,\rho_2$, i.e., $\Gamma(\rho_1,\rho_2) := \{\rho_{12} \in \Omega_{n^2}: \text{Tr}_1[\rho_{12}] = \rho_2, \text{Tr}_2[\rho_{12}] = \rho_1 \}$, where $\Tr_i$, $i=1,2$ denote the partial traces. Then the quantum $p$-Wasserstein semidistance is given by
\begin{align}
\label{wasserstein}
    W_p(\rho_1,\rho_2) = \min_{\rho_{12}\in\Gamma(\rho_1,\rho_2)} \left(\text{Tr}\left[\bE^p\rho_{12}\right]\right)^{1/p}, \quad \text{where} \quad \bE = \sum_{i < j} E_{ij} (\be_i \wedge \be_j)(\be_i \wedge \be_j)^\dagger,
\end{align}
with $(E_{ij})$ being some given distance matrix and the wedge product defined via $\bx \wedge \by \coloneqq \tfrac{1}{\sqrt{2}} (\bx \otimes \by - \by \otimes \bx) \in \Lambda^2\sC^n$. Thus, the operator $\bE$ on $\Lambda^2\sC^n$ can be interpreted as a cost matrix for a quantum analogue of the transport problem. If the states of interest are pure, $\rho_1 = \bx\bx^\dagger$, $\rho_2 = \by\by^\dagger$, then there exists only one coupling $\rho_{12} = \bx\bx^\dagger \otimes \by\by^\dagger$ \cite{CEFZ21}, and by explicit calculation one obtains that
\begin{align}
\label{wasserstein_pure}
    W_p(\bx\bx^\dagger,\by\by^\dagger) = \Big(\tfrac{1}{2}\sum\limits_{i<j} E^p_{ij}|x_iy_j - x_jy_i|^2 \Big)^{1/p}.
\end{align}

Notably, even though $W_p$ given by \eqref{wasserstein} is only a \emph{semi}-distance---for $n > 2$ it generally fails to satisfy the triangle inequality on the mixed states \cite{MillerComment}---numerical investigations strongly suggested that its restriction to pure states is a true distance. This conjecture was further supported by partial results gathered in \cite{llaproc}, where the triangle inequality on pure states was shown to hold in the $p=2$ case for various distance matrices $(E_{ij})$ (including the rank-one and Euclidean ones \cite[Theorems 6.4 \& 6.8]{llaproc}). However, the proof of the triangle inequality in full generality remained elusive.

The aim of the current work is to fill this gap and prove that $W_p$ on pure states is a bona fide distance for any $p \geq 2$ and any distance matrix $(E_{ij})$. For convenience, below we state the main results independently of the quantum optimal transport formulation above (and we also drop the cumbersome $\tfrac{1}{2}$ factor in front of the sum).

\smallskip

Fix an $n \times n$ distance matrix $(E_{ij})$ and consider the following map on $\PCn$ for any $p > 0$
\begin{align}
\label{dp}
d_p(\bx,\by) \vc \Big(\sum\limits_{i<j} E^p_{ij}|x_iy_j - x_jy_i|^2 \Big)^{1/p} = \|\bE^{p/2}(\bx \wedge \by)\|^{2/p},
\end{align}
where $\bE$ is a positive-definite operator on $\Lambda^2\sC^n$ defined via $\bE(\be_i \wedge \be_j) = E_{ij} \be_i \wedge \be_j$ (where $\be_1,\ldots,\be_n$ serve as a fixed computational basis), and the norm $\|.\|$ is induced by the standard inner product on $\Lambda^2\sC^n$ given by $\langle \bv_1 \wedge \bv_2 | \bw_1 \wedge \bw_2\rangle = \left| \begin{smallmatrix} \bv_1^\dag \bw_1 & \bv_1^\dag \bw_2 \\ \bv_2^\dag \bw_1 & \bv_2^\dag \bw_2 \end{smallmatrix} \right|$. Notice that formula \eqref{dp} significantly generalizes the Hilbert--Schmidt distance \eqref{eq:HSdist}. Indeed, by the Lagrange identity $d_{\text{HS}}(\bx,\by) = \|\bx \wedge \by\|$, and so $d_{\text{HS}}$ is a special case of $d_2$ involving the distance matrix $E_{ij} \vc 1 - \delta_{ij}$ (i.e., the distance matrix of the discrete metric). 

The map $d_p$ is clearly a well-defined semidistance (i.e., it is symmetric and nondegenerate) on $\PCn$, and the main result of this paper is demonstrating that \eqref{dp} respects also the triangle inequality for $p \geq 2$.
\begin{theorem}
\label{main}
Let $n \geq 2$. For any $n \times n$ distance matrix $(E_{ij})$ and any $p \geq 2$, the map $d_p: \PCn \x \PCn \rightarrow \sR$ given by \eqref{dp} is a distance function. In particular, it satisfies
\begin{align}
\label{triangle1}
d_p(\bx,\by) \leq d_p(\bx,\bz) + d_p(\by,\bz)
\end{align}
for any $\bx, \by, \bz \in \PCn$.
\end{theorem}

Several remarks are in order.
\begin{remark}
\label{rem1}
The case $n=2$ of Theorem \ref{main} is trivial, because one then has $d_p(\bx,\by) = E_{12} |x_1y_2 - x_2y_1|^{2/p} = E_{12} \|\bx \wedge \by\|^{2/p}$, which clearly defines a distance function, being a composition of the Hilbert--Schmidt distance with the increasing concave map $t \mapsto E_{12}t^{2/p}$.
\end{remark}

\begin{remark}
\label{rem_poltora}
For any $n \geq 2$, the map $d_p$ can be easily compared with the Hilbert--Schmidt distance. Indeed, notice that since the operator $\bE$ has eigenbasis $\{\be_i \wedge \be_j\}_{i<j}$ with $E_{ij}$ as the corresponding eigenvalues, then from \eqref{dp} it immediately follows that
\begin{align}
\label{bound1}
\min_{i<j} E_{ij} \, d^{2/p}_{\text{HS}}(\bx,\by) = \min_{i<j} E_{ij} \, \|\bx \wedge \by\|^{2/p} \leq d_p(\bx,\by) \leq \max_{i<j} E_{ij} \, \|\bx \wedge \by\|^{2/p} = \max_{i<j} E_{ij} \, d^{2/p}_{\text{HS}}(\bx,\by)
\end{align}
for any $\bx, \by \in \PCn$. In other words, $d_p$ is bi-H\"{o}lder continuous with respect to the Hilbert--Schmidt distance. Of course, since the latter is equivalent to the Fubini--Study distance (by the elementary inequality $2t/\pi \leq \sin t \leq t$ for all $t \in [0,\pi/2]$), one can also write that, for any $\bx, \by \in \PCn$,
\begin{align}
\label{bound2}
\left( \tfrac{2}{\pi} \right)^{2/p} \min_{i<j} E_{ij} \ d^{2/p}_{\text{FS}}(\bx,\by) \leq d_p(\bx,\by) \leq \max_{i<j} E_{ij} \ d^{2/p}_{\text{FS}}(\bx,\by).
\end{align}
\end{remark}

\begin{remark}
\label{rem2}
The assumption $p \geq 2$ in Theorem \ref{main} is necessary. More concretely, for $p < 2$ the map $d_p$ no longer satisfies the triangle inequality, regardless of the choice of $n$ and the distance matrix $(E_{ij})$.

Indeed, consider $\bx = \cos\theta \be_1 + \sin\theta \be_2$, $\by = \cos\theta \be_1 - \sin\theta \be_2$ and $\bz = \be_1$ (modulo phase factors) for some yet unspecified $\theta \in (0,\tfrac{\pi}{2})$. Then $d_p(\bx,\by) = E_{12} \sin^{2/p} 2\theta$, whereas $d_p(\bx,\bz) = d_p(\by,\bz) = E_{12} \sin^{2/p} \theta$, and so the triangle inequality would imply that $\sin 2\theta \leq 2^{p/2} \sin \theta$. However, this is false for any $\theta$ satisfying $\cos \theta > 2^{p/2-1}$, which exist within $(0,\tfrac{\pi}{2})$ precisely when $p < 2$.
\end{remark}

\begin{remark}
On superposition states, $d_p$ satisfies the following inequality
\begin{align*}
d_p(a \bx + b \by,\bz) \leq \left( |a| d^{p/2}_p(\bx,\bz) + |b| d^{p/2}_p(\by,\bz) \right)^{2/p}
\end{align*}
for any $\bx,\by,\bz \in \PCn$ with $\bx \bot \by$ and any $a,b \in \sC$, $|a|^2 + |b|^2 = 1$. It follows directly from \eqref{dp} by the subadditivity of $\|.\|$.
\end{remark}

\begin{figure}[h!]
\centering
\begin{tikzpicture}[x=0.75pt,y=0.75pt,yscale=-1,xscale=1]

\draw  [draw opacity=0] (296.59,51.81) .. controls (324.31,37.02) and (357.98,41.04) .. (377.97,63.61) .. controls (401.73,90.44) and (397.13,133.33) .. (367.69,159.41) .. controls (338.25,185.48) and (295.12,184.87) .. (271.35,158.03) .. controls (258.34,143.35) and (253.83,123.85) .. (257.24,104.94) -- (324.66,110.82) -- cycle ; \draw   (296.59,51.81) .. controls (324.31,37.02) and (357.98,41.04) .. (377.97,63.61) .. controls (401.73,90.44) and (397.13,133.33) .. (367.69,159.41) .. controls (338.25,185.48) and (295.12,184.87) .. (271.35,158.03) .. controls (258.34,143.35) and (253.83,123.85) .. (257.24,104.94) ;  
\draw  [draw opacity=0] (274.9,110.01) .. controls (266.22,109.47) and (257.82,106.34) .. (250.86,100.48) .. controls (232.45,84.96) and (231.19,56.16) .. (248.05,36.15) .. controls (264.91,16.14) and (293.51,12.5) .. (311.93,28.01) .. controls (317.05,32.33) and (320.84,37.66) .. (323.29,43.55) -- (281.4,64.25) -- cycle ; \draw   (274.9,110.01) .. controls (266.22,109.47) and (257.82,106.34) .. (250.86,100.48) .. controls (232.45,84.96) and (231.19,56.16) .. (248.05,36.15) .. controls (264.91,16.14) and (293.51,12.5) .. (311.93,28.01) .. controls (317.05,32.33) and (320.84,37.66) .. (323.29,43.55) ;  
\draw  [draw opacity=0][shading=_g1pc9w3t7,_4m3ns55d7] (340.23,71.6) .. controls (340.23,70.16) and (341.41,68.99) .. (342.85,68.99) .. controls (344.29,68.99) and (345.47,70.16) .. (345.47,71.6) .. controls (345.47,73.05) and (344.29,74.22) .. (342.85,74.22) .. controls (341.41,74.22) and (340.23,73.05) .. (340.23,71.6) -- cycle ;
\draw  [draw opacity=0][shading=_qbtkb5o37,_jgeqz7du0] (365.23,122.56) .. controls (365.23,121.12) and (366.4,119.95) .. (367.85,119.95) .. controls (369.29,119.95) and (370.46,121.12) .. (370.46,122.56) .. controls (370.46,124.01) and (369.29,125.18) .. (367.85,125.18) .. controls (366.4,125.18) and (365.23,124.01) .. (365.23,122.56) -- cycle ;
\draw  [draw opacity=0][shading=_opmlqqfi0,_nt16o215o] (241.69,64.52) .. controls (241.69,63.07) and (242.86,61.9) .. (244.3,61.9) .. controls (245.75,61.9) and (246.92,63.07) .. (246.92,64.52) .. controls (246.92,65.96) and (245.75,67.13) .. (244.3,67.13) .. controls (242.86,67.13) and (241.69,65.96) .. (241.69,64.52) -- cycle ;
\draw  [draw opacity=0][shading=_u01x1z4a0,_yumybsto8] (288.6,24.69) .. controls (288.6,23.25) and (289.77,22.08) .. (291.21,22.08) .. controls (292.66,22.08) and (293.83,23.25) .. (293.83,24.69) .. controls (293.83,26.14) and (292.66,27.31) .. (291.21,27.31) .. controls (289.77,27.31) and (288.6,26.14) .. (288.6,24.69) -- cycle ;
\draw  [draw opacity=0][shading=_itvxi5kv9,_9uy8x5mc0] (292.99,167.11) .. controls (292.99,165.67) and (294.16,164.5) .. (295.6,164.5) .. controls (297.05,164.5) and (298.22,165.67) .. (298.22,167.11) .. controls (298.22,168.56) and (297.05,169.73) .. (295.6,169.73) .. controls (294.16,169.73) and (292.99,168.56) .. (292.99,167.11) -- cycle ;
\draw  [draw opacity=0][shading=_3nmh2rxaf,_yw3sbxir5] (103.89,71.6) .. controls (103.89,70.16) and (105.06,68.99) .. (106.5,68.99) .. controls (107.95,68.99) and (109.12,70.16) .. (109.12,71.6) .. controls (109.12,73.05) and (107.95,74.22) .. (106.5,74.22) .. controls (105.06,74.22) and (103.89,73.05) .. (103.89,71.6) -- cycle ;
\draw  [draw opacity=0][shading=_fh5lwj2sk,_03k4z8nuo] (128.89,122.56) .. controls (128.89,121.12) and (130.06,119.95) .. (131.5,119.95) .. controls (132.95,119.95) and (134.12,121.12) .. (134.12,122.56) .. controls (134.12,124.01) and (132.95,125.18) .. (131.5,125.18) .. controls (130.06,125.18) and (128.89,124.01) .. (128.89,122.56) -- cycle ;
\draw  [draw opacity=0][shading=_fpftkbl0x,_8fgmug454] (5.34,64.52) .. controls (5.34,63.07) and (6.51,61.9) .. (7.96,61.9) .. controls (9.4,61.9) and (10.57,63.07) .. (10.57,64.52) .. controls (10.57,65.96) and (9.4,67.13) .. (7.96,67.13) .. controls (6.51,67.13) and (5.34,65.96) .. (5.34,64.52) -- cycle ;
\draw  [draw opacity=0][shading=_4mn34k8wq,_z0xc2egj0] (52.25,24.69) .. controls (52.25,23.25) and (53.42,22.08) .. (54.87,22.08) .. controls (56.31,22.08) and (57.48,23.25) .. (57.48,24.69) .. controls (57.48,26.14) and (56.31,27.31) .. (54.87,27.31) .. controls (53.42,27.31) and (52.25,26.14) .. (52.25,24.69) -- cycle ;
\draw  [draw opacity=0][shading=_20zjeydf3,_jv66gayjb] (56.64,167.11) .. controls (56.64,165.67) and (57.81,164.5) .. (59.26,164.5) .. controls (60.7,164.5) and (61.87,165.67) .. (61.87,167.11) .. controls (61.87,168.56) and (60.7,169.73) .. (59.26,169.73) .. controls (57.81,169.73) and (56.64,168.56) .. (56.64,167.11) -- cycle ;
\draw    (172.5,107.12) -- (172.84,107.12) -- (218.09,107.12) ;
\draw [shift={(220.09,107.12)}, rotate = 180] [fill={rgb, 255:red, 0; green, 0; blue, 0 }  ][line width=0.08]  [draw opacity=0] (12,-3) -- (0,0) -- (12,3) -- cycle    ;
\draw [shift={(172.5,107.12)}, rotate = 0] [color={rgb, 255:red, 0; green, 0; blue, 0 }  ][line width=0.75]      (0,-11.18) .. controls (-3.09,-11.18) and (-5.59,-8.68) .. (-5.59,-5.59) .. controls (-5.59,-2.5) and (-3.09,0) .. (0,0) ;
\draw  [draw opacity=0][shading=_kgttk51lb,_1igd3srsx] (288.26,81.39) .. controls (288.26,79.95) and (289.43,78.78) .. (290.88,78.78) .. controls (292.32,78.78) and (293.49,79.95) .. (293.49,81.39) .. controls (293.49,82.84) and (292.32,84.01) .. (290.88,84.01) .. controls (289.43,84.01) and (288.26,82.84) .. (288.26,81.39) -- cycle ;
\draw  [draw opacity=0][shading=_dym6cyghs,_nsi30h14y] (52.02,81.39) .. controls (52.02,79.95) and (53.19,78.78) .. (54.64,78.78) .. controls (56.08,78.78) and (57.25,79.95) .. (57.25,81.39) .. controls (57.25,82.84) and (56.08,84.01) .. (54.64,84.01) .. controls (53.19,84.01) and (52.02,82.84) .. (52.02,81.39) -- cycle ;

\draw (8.96,67.92) node [anchor=north west][inner sep=0.75pt]  [font=\normalsize]  {${\displaystyle a_{1}}$};
\draw (55.87,28.09) node [anchor=north west][inner sep=0.75pt]  [font=\normalsize]  {${\displaystyle \mathnormal{a}_{2}}$};
\draw (107.5,75) node [anchor=north west][inner sep=0.75pt]  [font=\normalsize]  {${\displaystyle a_{4}}$};
\draw (60.26,170.51) node [anchor=north west][inner sep=0.75pt]  [font=\normalsize]  {${\displaystyle a_{5}}$};
\draw (132.5,125.96) node [anchor=north west][inner sep=0.75pt]  [font=\normalsize]  {${\displaystyle a_{6}}$};
\draw (245.3,67.92) node [anchor=north west][inner sep=0.75pt]  [font=\normalsize]  {${\displaystyle \mathbf{e}_{1}}$};
\draw (292.21,28.09) node [anchor=north west][inner sep=0.75pt]  [font=\normalsize]  {${\displaystyle \mathbf{e}_{2}}$};
\draw (343.85,75) node [anchor=north west][inner sep=0.75pt]  [font=\normalsize]  {${\displaystyle \mathbf{e}_{4}}$};
\draw (368.85,125.96) node [anchor=north west][inner sep=0.75pt]  [font=\normalsize]  {${\displaystyle \mathbf{e}_{6}}$};
\draw (299.91,161.81) node [anchor=north west][inner sep=0.75pt]  [font=\normalsize]  {${\displaystyle \mathbf{e}_{5}}$};
\draw (185.74,86.76) node [anchor=north west][inner sep=0.75pt]  [font=\large]  {$\iota $};
\draw (41.18,190.46) node [anchor=north west][inner sep=0.75pt]  [font=\normalsize]  {$( \cX,\dist )$};
\draw (275.79,188.35) node [anchor=north west][inner sep=0.75pt]  [font=\normalsize]  {$\left(\mathbb{P}\left(\mathbb{C}^{6}\right) ,d_{p}\right)$};
\draw (291.88,84.79) node [anchor=north west][inner sep=0.75pt]  [font=\normalsize]  {${\displaystyle \mathbf{e}_{3}}$};
\draw (55.64,84.79) node [anchor=north west][inner sep=0.75pt]  [font=\normalsize]  {${\displaystyle \mathnormal{a}_{3}}$};
\end{tikzpicture}
\caption{An illustration of Remark \ref{remint} (for $n=6$). Given any $n$-element metric space $(\cX,\dist)$, we can isometrically embed it into $(\PCn,d_p)$ with the distance function $d_p$ defined via \eqref{dp} for any chosen $p \geq 2$. In doing so, the elements $a_i$ of $\cX$ are promoted to the orthonormal basis states $\be_i$ (defined up to a phase factor). Loosely speaking, the above construction provides a general way to `quantize' any finite metric space.}
\label{figurka}
\end{figure}
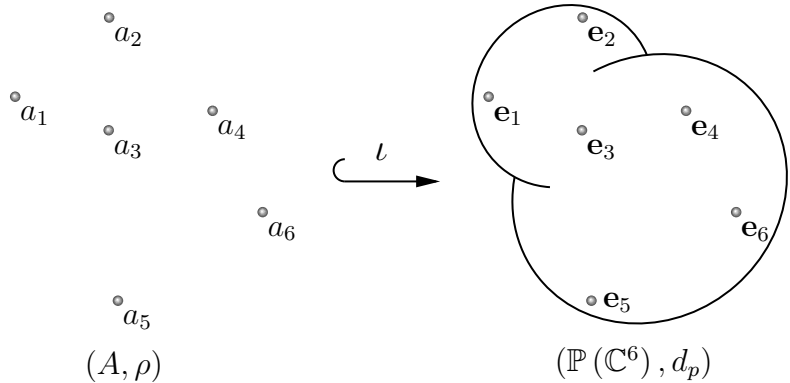

\begin{remark}
\label{remint}
The result that $d_p$ is a true distance function can be seen as providing a way to isometrically embed any $n$-element metric space into the space of pure quantum states `spanned' upon its elements (cf. Figure \ref{figurka}). More concretely, given a finite metric space $(\cX,\dist)$ with $\cX \vc \{a_1,\ldots,a_n\}$ and the associated distance matrix $E_{ij} \vc \dist(a_i,a_j)$, the map $\iota: \cX \rightarrow \PCn$ defined via $\iota(a_i) \vc \be_i$ (modulo phase factor) is an \emph{isometric embedding} of $(\cX,\dist)$ into $(\PCn,d_p)$, because clearly 
\begin{align*}
d_p(\iota(a_i),\iota(a_j)) = d_p(\be_i,\be_j) = E_{ij} = \dist(a_i,a_j).
\end{align*}
\end{remark}

\begin{remark}
\label{rem3}
Although the elements of $\PCn$, when regarded as norm-one vectors from $\sC^n$, are defined up to a phase factor, we can---and will---forget about the latter cumbersome fact from now on. In particular, we shall prove the triangle inequality \eqref{triangle1} for any $\bx, \by, \bz \in \sC^n$ such that $\|\bx\| = \|\by\| = \|\bz\| = 1$.
\end{remark}

Before we delve into the proof of Theorem \ref{main}, let us address the natural question concerning the symmetries of the proposed distances, i.e., which projective unitary transformations of $\PCn$ leave $d_p$ invariant. Clearly, the answer strongly depends on the underlying distance matrix $(E_{ij})$ and its associated operator $\bE$. More concretely, we have the following result.
\begin{proposition}
\label{prop_sym}
For any $\bU \in \textup{U}(n)$ the following conditions are equivalent
\begin{enumerate}[(i)]
\item $d_p(\bU\bx,\bU\by) = d_p(\bx,\by)$ for any $\bx,\by \in \PCn$.
\item $\bE$ commutes with $\bU \wedge \bU$.
\end{enumerate}
\end{proposition}
\begin{proof}
Let us define the quadratic form $q$ on $\Lambda^2\sC^n$ via 
\begin{align}
\label{lem_q}
q(\bV) \vc \| \bE^{p/2}(\bU \wedge \bU)\bV \|^2 = \langle \bV | (\bU^\dag \wedge \bU^\dag) \bE^p (\bU \wedge \bU) \bV \rangle
\end{align}
for any $\bV \in \Lambda^2\sC^n$. Notice that, by \eqref{dp}, condition \emph{(i)} can be now restated simply as
\begin{align}
\label{lem_i}
\forall \bx,\by \in \PCn \quad q(\bx \wedge \by) = \| \bE^{p/2} (\bx \wedge \by) \|^2.
\end{align}

In order to prove \emph{(i)} $\Rightarrow$ \emph{(ii)}, it suffices to show that, for any indices $i,j,k,l$,
\begin{align}
\label{lem_ii_prim}
\langle \be_i \wedge \be_j | (\bU^\dag \wedge \bU^\dag) \bE^p (\bU \wedge \bU) (\be_k \wedge \be_l) \rangle = \langle \be_i \wedge \be_j | \bE^p (\be_k \wedge \be_l) \rangle,
\end{align}
where $\{\be_i \wedge \be_j\}_{i<j}$ is the eigenbasis of $\bE$ and so the right-hand side is in fact nothing but $E_{ij}^p(\delta_{ik}\delta_{jl}-\delta_{il}\delta_{jk})$. Indeed, \eqref{lem_ii_prim} would give the operator equality $(\bU^\dag \wedge \bU^\dag) \bE^p (\bU \wedge \bU) = \bE^p$, which by the unitarity of $\bU$ could be then transformed to $\bE (\bU \wedge \bU) = (\bU \wedge \bU) \bE$.

It is convenient to split the proof of \eqref{lem_ii_prim} into three cases.

For the case when the pairs of indices $\{i,j\}$ and $\{k,l\}$ coincide, one obtains \eqref{lem_ii_prim} directly from \eqref{lem_i} simply by taking $\bx := \be_i$ and $\by := \be_j$.

For the case when the pairs $\{i,j\}$ and $\{k,l\}$ share exactly one index, say $i = l$, substituting $\bx := \be_i$ and $\by := \tfrac{1}{\sqrt{1+|z|^2}} \left(\be_j + z \be_k \right)$ for any fixed $z \in \sC$ into condition \eqref{lem_i} gives
\begin{align*}
& q(\be_i \wedge \be_j) + |z|^2 q(\be_i \wedge \be_k) + 2\RE\left[ z \langle \be_i \wedge \be_j | (\bU^\dag \wedge \bU^\dag) \bE^p (\bU \wedge \bU) (\be_i \wedge \be_k) \rangle \right] = E_{ij}^p + |z|^2 E_{ik}^p.
\end{align*}
But since $q(\be_i \wedge \be_j) = E_{ij}^p$ and $q(\be_i \wedge \be_k) = E_{ik}^p$ by \eqref{lem_i}, the above equality reduces to the vanishing of the remaining term, which in turn implies, by the arbitrariness of $z$, that
\begin{align}
\label{lem_case2}
\langle \be_i \wedge \be_j | (\bU^\dag \wedge \bU^\dag) \bE^p (\bU \wedge \bU) (\be_k \wedge \be_i) \rangle = 0,
\end{align}
which is nothing but \eqref{lem_ii_prim} in the considered case.

Finally, for the case when all indices $i,j,k,l$ are distinct, let us substitute $\bx := \tfrac{1}{\sqrt{1+|z|^2}} \left(\be_i + z \be_k \right)$ and $\by := \tfrac{1}{\sqrt{1+|\zeta|^2}} \left(\be_j + \zeta \be_l \right)$ for any fixed $z,\zeta \in \sC$ into condition \eqref{lem_i}. Similarly as before, the quadratic terms on both sides cancel out, whereas by \eqref{lem_case2} many cross terms vanish and one is left with the equality
\begin{align*}
\RE \left[ z \zeta \langle \be_i \wedge \be_j | (\bU^\dag \wedge \bU^\dag) \bE^p (\bU \wedge \bU) (\be_k \wedge \be_l) \rangle + \bar{z} \zeta \langle \be_k \wedge \be_j | (\bU^\dag \wedge \bU^\dag) \bE^p (\bU \wedge \bU) (\be_i \wedge \be_l) \rangle \right] = 0.
\end{align*}
By the arbitrariness of $z$ and $\zeta$, both above cross terms must vanish, thus proving \eqref{lem_ii_prim} in the final case.

To prove the converse implication \emph{(ii)} $\Rightarrow$ \emph{(i)}, notice that the commutation of $\bE$ and $\bU \wedge \bU$ simplifies formula \eqref{lem_q} defining $q$ to
\begin{align*}
\forall \bV \in \Lambda^2\sC^n \quad q(\bV) = \| \bE^{p/2} \bV \|^2,
\end{align*}
and hence \emph{(i)} follows trivially.
\end{proof}

\begin{remark}
\label{rem_sym}
The subgroup of unitaries $\bU$ such that $[\bE, \bU \wedge \bU] = 0$ can vary depending on the details of the distance matrix $(E_{ij})$. For example, if all $E_{ij}$'s are equal (for $i \neq j$), then $\bE$ is a scalar operator and as such it commutes with $\bU \wedge \bU$ for all $\bU \in \textup{U}(n)$ (which is not surprising, because in that case $d_p \propto d_\textup{HS}^{2/p}$). On the other hand, as another extreme example consider the case where all $E_{ij}$'s are \emph{distinct} (for $i < j$ with $n \geq 3$). Then the commutation relation $[\bE, \bU \wedge \bU] = 0$ forces $\bU \wedge \bU$ to be diagonal in the basis $\{\be_i \wedge \be_j\}_{i<j}$, which in turn forces $\bU$ to be diagonal in the basis $\{\be_i\}$. These are thus the only $\bU \in \textup{U}(n)$ leaving $d_p$ invariant in this case.
\end{remark}

Theorem \ref{main} has further implications and raises several natural questions for future investigation.

First, it provides a firm theoretical foundation for the use of the distances $d_p$ in rigorous applications throughout quantum information science. Moreover, in light of Remark \ref{rem2}, one may argue that \eqref{dp} provides a canonical way to endow $\PCn$ with a metric-space structure induced by a given distance matrix. Finally, since $d_p$ is, up to a factor of $2^{-1/p}$, the restriction of the quantum $p$-Wasserstein semidistance \eqref{wasserstein} to pure states, Theorem \ref{main} provides further support for, and an illustration of, the formulation of the quantum transport problem proposed in \cite{FECZ21}.

The distances $d_p$ may be useful in physical settings involving pure states in which the geometry of the system, encoded by the distance matrix, must be taken into account. As a first example, consider an energy-like distance associated with a Hermitian matrix $H$, interpreted as the Hamiltonian. Take the computational basis to be an eigenbasis of $H$, and define the entries of the distance matrix by the corresponding spectral gaps, $E_{ij} = |\lambda_i- \lambda_j|$. The properties and applications of $d_p$ induced by this distance matrix are discussed in the forthcoming paper \cite{Energy_distance}.

Another example arises from an $m$-qubit system, whose pure-state space is $\mathbb{P}(\mathbb{C}^{2^m})$. Each vector of the computational basis can be identified with a bit string of length $m$, and the entries $E_{ij}$ of the distance matrix can be chosen to be the Hamming distances between the corresponding bit strings. The induced distance $d_p$ then provides a concise way of defining a ``quantum $p$-Hamming distance'' between arbitrary pure states of the system, including highly entangled states. Such a distance may be useful in problems in quantum machine learning, where Hamming-based divergences are used to distinguish multiqubit quantum states \cite{U_learn,Hamming_distance}. Note that this approach to `quantizing' the Hamming distance differs from that proposed by De Palma et al. in \cite{PMTL21}.

Finally, one might ask whether the distances $d_p$ can be extended to mixed states while retaining all the metric axioms, including the triangle inequality. In this paper we provide a nontrivial, albeit partial, answer by proposing the concise formula \eqref{dhat}, which defines a genuine metric on the set of density matrices for a broad class of distance matrices.

The outline of the paper is as follows. In section \ref{sec::2} we prove Theorem \ref{main} for the first nontrivial case $n=3$ (Proposition \ref{n3case}) and then show that, for $n>3$, it suffices to prove \eqref{triangle1} for $\{\bx,\by,\bz\}$ \emph{orthonormal} (Proposition \ref{partialresult}). The latter is done in section \ref{sec::3} by first establishing an auxiliary convexity result (Theorem \ref{convexThm}). Finally, in section \ref{sec::4} we propose and study an extension of the distances $d_p$ to mixed states, which unlike the original optimal-transport formulation defines genuine distances between density matrices.

\section{The $n=3$ case and its consequences}
\label{sec::2}

Just as announced in Remark \ref{rem3}, in what follows we will regard $d_p$ as defined on the (pairs of) norm-one vectors, and not their equivalence classes modulo phase factor. For $n=3$, this in particular means that definition \eqref{dp} can be written with the help of the cross product as $d_p(\bx,\by) = \|\bE^{p/2} (\bx \times \by)\|^{2/p}$, where $\bE = \diag(E_{23},E_{13},E_{12})$. 

For later use, let us recall some basic identities involving the cross product (valid for any $\sC^3$ vectors, not just those of norm 1):
\begin{align}
\label{id1}
& \bx^\top(\by \x \bz) = \by^\top(\bz \x \bx) = \bz^\top(\bx \x \by)
\\
\label{id2}
& (\bx \x \by)^\top(\bz \x \bu) = (\bx^\top\bz)(\by^\top\bu)-(\bx^\top\bu)(\by^\top\bz)
\\
\label{id3}
& \bx \x (\by \x \bz) = (\bx^\top\bz)\by - (\bx^\top\by)\bz
\\
\label{id4}
& (\bx \x \bz) \x (\by \x \bz) = (\bz^\top(\bx \x \by))\bz
\\
\label{id5}
& \bA\bx \x \bA\by = (\cof \bA)(\bx \x \by), \qquad \bA \in \sM_3(\sC)
\end{align}
where $\cof \bA$ denotes the cofactor matrix of $\bA$. Identity \eqref{id2} in particular implies that\footnote{Identity \eqref{id6} can also be viewed as the special case of the Lagrange identity.}
\begin{align}
\label{id6}
& \|\bx \x \by\|^2 = \|\bx\|^2\|\by\|^2-|\bx^\dag\by|^2,
\end{align}
which in turn implies that $\bx \x \by = 0$ if and only if $\bx$ and $\by$ are linearly dependent.

The following simple fact will also turn out to be important,
\begin{align}
\label{id7}
& \RE(\bx^\dag\by) = \RE \bx^\top \RE \by + \IM \bx^\top \IM \by.
\end{align}

The main result of this section is as follows.
\begin{proposition}
\label{n3case}
Fix any $p \geq 2$ and any positive-definite $\bE \in \sM_3(\sC)$. The triangle inequality
\begin{align}
\label{ineq3}
\|\bE^{p/2}(\bx \times \by)\|^{2/p} \leq \|\bE^{p/2}(\bx \times \bz)\|^{2/p} + \|\bE^{p/2}(\by \times \bz)\|^{2/p}, \quad \|\bx\| = \|\by\| = \|\bz\| = 1
\end{align}
holds if and only if the spectral radius $\rho(\bE)$ satisfies\footnote{For $\bE$ positive-definite, condition \eqref{war} is equivalent to the triangle inequality for its eigenvalues. Note that instead of $\rho(\bE)$ one can equally well use the spectral norm $\|\bE\|_{\textup{sp}}$ or the numerical radius $r(\bE)$.}
\begin{align}
\label{war}
2 \rho(\bE) \leq \Tr \bE.
\end{align}
\end{proposition}
\begin{proof}
Without loss of generality, we may regard $\bE$ as a diagonal matrix, because for any unitary matrix $\bU$, on the strength of \eqref{id5},
\begin{align*}
\|\bE^{p/2}(\bx \x \by)\| = \|\bU^\dag\bE^{p/2}\bU\bU^\dag(\bx \x \by)\| = \|(\bU^\dag\bE\bU)^{p/2}(\bU^\top\bx \x \bU^\top\by)\|,
\end{align*}
so $\bE$ can always be diagonalized. Of course, condition \eqref{war} is invariant under unitary rotations.

From now on we thus assume that $\bE = \textup{diag}(\lambda_1,\lambda_2,\lambda_3)$ (with $\lambda_i > 0$). Verifying the necessity of condition \eqref{war} is easy. Plugging the canonical basis $\bx = \be_1$, $\by = \be_2$, $\bz = \be_3$ into \eqref{ineq3} yields $\lambda_3 \leq \lambda_2 + \lambda_1$ and permuting the indices gives the other two inequalities. Altogether, this yields $2\max_i \lambda_i \leq \lambda_1 + \lambda_2 + \lambda_3$, which is nothing but \eqref{war}.

Proving the sufficiency of \eqref{war}, however, is much more complicated. The idea is to minimize the function $\|\bE^{p/2}(\bx \x \bz)\|^{2/p} + \|\bE^{p/2}(\bz \x \by)\|^{2/p} - \|\bE^{p/2}(\bx \x \by)\|^{2/p}$ subject to the constraints $\|\bx\| = \|\by\| = \|\bz\| = 1$ using the method of Lagrange multipliers.

Concretely, let us denote $\bM \vc \bE^{p/2}$ and consider the function $L: \sC^3 \x \sC^3 \x \sC^3 \x \sR^3 \cong \sR^{21} \rightarrow \sR$ defined as
\begin{align*}
L(\bx,\by,\bz; \alpha, \beta, \gamma) & = \|\bM(\bx \x \bz)\|^{2/p} + \|\bM(\by \x \bz)\|^{2/p} - \|\bM(\bx \x \by)\|^{2/p}
\\
& - \tfrac{1}{p}\alpha(\|\bx\|^2-1) - \tfrac{1}{p}\beta(\|\by\|^2-1) - \tfrac{1}{p}\gamma(\|\bz\|^2-1).
\end{align*}
We are interested in finding its stationary points as well as points of non-differentiability, because only at those points can $L$ attain global minimum (subject to our constraints).

The points of non-differentiability belonging to the feasible set are exactly those for which either
\begin{itemize}
\item $\bx \x \bz = 0$, but then for $\|\bx\| = \|\bz\| = 1$ we must have $\bz = \ee^{\ii\phi}\bx$ and so \eqref{ineq3} holds as an equality;
\item $\by \x \bz = 0$, but then we similarly obtain that $\bz = \ee^{\ii\phi}\by$ and \eqref{ineq3} holds as an equality;
\item $\bx \x \by = 0$, but then the LHS in \eqref{ineq3} is zero.
\end{itemize}

Having taken care of the points of non-differentiability, we can from now on assume that $\bx \x \by$, $\bx \x \bz$ and $\by \x \bz$ are all nonzero and calculate the total derivative $\dd L$ term by term. Denoting $M_{13} \vc \|\bM(\bx \x \bz)\|$, one obtains, for instance,
\begin{align*}
\dd\|\bM(\bx \x \bz)\|^{2/p} & = \frac{1}{p M_{13}^{2(p-1)/p}} \dd((\bx \x \bz)^\dag\bM^2(\bx \x \bz))
\\
& = \frac{2}{p M_{13}^{2(p-1)/p}} \RE\left[(\bM^2(\bx \x \bz))^\dag(\dd\bx \x \bz + \bx \x \dd\bz)\right]
\\
& = \frac{2}{p M_{13}^{2(p-1)/p}} \Big( \RE\Big[(\bar{\bz} \x \bM^2(\bx \x \bz))^\dag\dd\bx\Big]- \RE\Big[(\bar{\bx} \x \bM^2(\bx \x \bz))^\dag\dd\bz \Big]\Big),
\end{align*}
where in the last equality we have used \eqref{id1}. Calculating $\dd\|\bM(\by \x \bz)\|^{2/p}$ and $\dd\|\bM(\bx \x \by)\|^{2/p}$ similarly, along with the total derivatives of the terms involving the Lagrange multipliers, such as
\begin{align*}
\dd \left(\tfrac{1}{p}\alpha(\|\bx\|^2-1)\right) = \tfrac{2}{p}\alpha \RE (\bx^\dag\dd\bx) + \tfrac{1}{p}(\|\bx\|^2-1)\dd\alpha,
\end{align*}
one arrives at $\textup{d}L$ of the following form
\begin{align*}
\frac{p}{2} \dd L & = \RE \Bigg[ \left( \frac{\bar{\bz} \x \bM^2(\bx \x \bz)}{M_{13}^{2(p-1)/p}} - \frac{\bar{\by} \x \bM^2(\bx \x \by)}{M_{12}^{2(p-1)/p}} - \alpha \bx \right)^\dag\dd\bx \Bigg]
\\
& + \RE {\Bigg[} \left( \frac{\bar{\bz} \x \bM^2(\by \x \bz)}{M_{23}^{2(p-1)/p}} + \frac{\bar{\bx} \x \bM^2(\bx \x \by)}{M_{12}^{2(p-1)/p}} - \beta \by \right)^\dag\dd\by \Bigg]
\\
& + \RE  \Bigg[\left( - \frac{\bar{\bx} \x \bM^2(\bx \x \bz)}{M_{13}^{2(p-1)/p}} - \frac{\bar{\by} \x \bM^2(\by \x \bz)}{M_{23}^{2(p-1)/p}} - \gamma \bz \right)^\dag\dd\bz \Bigg]
\\
& - \tfrac{1}{2}(\|\bx\|^2-1)\dd\alpha - \tfrac{1}{2}(\|\by\|^2-1)\dd\beta - \tfrac{1}{2}(\|\bz\|^2-1)\dd\gamma,
\end{align*}
where we have denoted $M_{12} \vc \|\bM(\bx \x \by)\|$, $M_{13} \vc \|\bM(\bx \x \bz)\|$ and $M_{23} \vc \|\bM(\by \x \bz)\|$.

The method of Lagrange multipliers works for functions of \emph{real} variables and formally $L$ is a function of the \emph{six real 3-vectors}: $\RE\bx, \IM\bx, \RE\by, \IM\by, \RE\bz, \IM\bz$ (along with $\alpha$, $\beta$ and $\gamma$). Luckily, if we now apply \eqref{id7}, the differentials of these real 3-vectors (i.e. $\dd\RE\bx = \RE\dd\bx$, $\dd\IM\bx = \IM\dd\bx$, etc.) all explicitly appear next to the real and imaginary parts of the complex 3-vectors in the large parentheses above. Vanishing of all the first partial derivatives can be thus expressed as the following system of three complex vector equations + the three original constraints:
\begin{align}
\label{eqs1}
& \left\{ \begin{array}{l}
\alpha \bx = \frac{\bar{\bz} \x \bM^2(\bx \x \bz)}{M_{13}^{2(p-1)/p}} - \frac{\bar{\by} \x \bM^2(\bx \x \by)}{M_{12}^{2(p-1)/p}}
\\[8pt]
\beta \by = \frac{\bar{\bz} \x \bM^2(\by \x \bz)}{M_{23}^{2(p-1)/p}} + \frac{\bar{\bx} \x \bM^2(\bx \x \by)}{M_{12}^{2(p-1)/p}}
\\[8pt]
\gamma \bz = - \frac{\bar{\bx} \x \bM^2(\bx \x \bz)}{M_{13}^{2(p-1)/p}} - \frac{\bar{\by} \x \bM^2(\by \x \bz)}{M_{23}^{2(p-1)/p}}
\\[8pt]
\|\bx\| = \|\by\| = \|\bz\| = 1
\end{array} \right.
\end{align}

In order to extract some information from equations \eqref{eqs1}, let us take their inner products with $\bx$, $\by$, $\bz$, obtaining
\begin{align}
\label{eqs2}
\alpha = M_{13}^{2/p} - M_{12}^{2/p}, \quad \beta = M_{23}^{2/p} - M_{12}^{2/p}, \quad \gamma = M_{13}^{2/p} + M_{23}^{2/p}
\end{align}
along with
\begin{align*}
& \alpha \by^\dag\bx = \frac{(\by \x \bz)^\dag \bM^2 (\bx \x \bz)}{M_{13}^{2(p-1)/p}}, & \beta \bx^\dag\by = \frac{(\bx \x \bz)^\dag \bM^2 (\by \x \bz)}{M_{23}^{2(p-1)/p}},
\\
& \gamma \bx^\dag\bz = - \frac{(\bx \x \by)^\dag \bM^2 (\by \x \bz)}{M_{23}^{2(p-1)/p}}, & \alpha \bz^\dag\bx = \frac{(\by \x \bz)^\dag \bM^2 (\bx \x \by)}{M_{12}^{2(p-1)/p}},
\\
& \beta \bz^\dag\by = - \frac{(\bx \x \bz)^\dag \bM^2 (\bx \x \by)}{M_{12}^{2(p-1)/p}}, & \gamma \by^\dag\bz = \frac{(\bx \x \by)^\dag \bM^2 (\bx \x \bz)}{M_{13}^{2(p-1)/p}}.
\end{align*}
We have the following implications.
\begin{itemize}
\item If $\bx^\dag\by \neq 0$, then $\alpha M_{13}^{2(p-1)/p} = \beta M_{23}^{2(p-1)/p}$, which by \eqref{eqs2} leads to 
\begin{align*}
M_{23}^2 - M_{13}^2 = \left(M_{23}^{2(p-1)/p} - M_{13}^{2(p-1)/p}\right) M_{12}^{2/p},
\end{align*}
which in turn implies $M_{13} = M_{23}$ \textbf{or} $M_{12}^{2/p} = \frac{M_{13}^2 - M_{23}^2}{M_{13}^{2(p-1)/p} - M_{23}^{2(p-1)/p}}$. 

In the latter case, the triangle inequality \eqref{ineq3}, in the current notation expressed simply as $M_{12}^{2/p} \leq M_{13}^{2/p} + M_{23}^{2/p}$, is satisfied because
\begin{align*}
M_{12}^{2/p} = \frac{M_{13}^2 - M_{23}^2}{M_{13}^{2(p-1)/p} - M_{23}^{2(p-1)/p}} \leq M_{13}^{2/p} + M_{23}^{2/p},
\end{align*}
where the inequality follows\footnote{To see how, divide both sides of the above inequality by $M_{13}^{2/p}$ or $M_{23}^{2/p}$, depending on which is larger.} from $\frac{a^p - 1}{a^{p-1} - 1} \leq a+1$, valid for any $a>1$ and $p \geq 2$.
\item If $\bx^\dag\bz \neq 0$, then $\alpha M_{12}^{2(p-1)/p} = -\gamma M_{23}^{2(p-1)/p}$, which on the strength of \eqref{eqs2} leads to $M_{13}^{2/p} = \frac{M_{12}^2 - M_{23}^2}{M_{12}^{2(p-1)/p} + M_{23}^{2(p-1)/p}}$. But the triangle inequality \eqref{ineq3} is then satisfied, because:
\begin{itemize}
\item For $M_{12} \geq M_{23}$ one can estimate (using $p \geq 2$ in the process)
\begin{align*}
M_{13}^{2/p} = \frac{M_{12}^2 - M_{23}^2}{M_{12}^{2(p-1)/p} + M_{23}^{2(p-1)/p}} \geq M_{12}^{2/p} - M_{23}^{2/p}.
\end{align*}
\item For $M_{12} < M_{23}$ one trivially has $M_{12}^{2/p} \leq M_{13}^{2/p} + M_{23}^{2/p}$, too.
\end{itemize}
\item If $\by^\dag\bz \neq 0$, then $\beta M_{12}^{2(p-1)/p} = -\gamma M_{13}^{2(p-1)/p}$, which on the strength of \eqref{eqs2} leads to $M_{23}^{2/p} = \frac{M_{12}^2 - M_{13}^2}{M_{12}^{2(p-1)/p} + M_{13}^{2(p-1)/p}}$. Similarly as in the previous case, the triangle inequality \eqref{ineq3} is then satisfied---simply swap $M_{23}$ with $M_{13}$ in the above reasoning.
\end{itemize}

Thus, only two cases remain:
\begin{enumerate}[1${}^\circ$]
\item $\{\bx, \by, \bz\}$ is an orthonormal system.
\item $\bx^\dag\bz = \by^\dag\bz = 0$ and $\|\bM(\bx \x \bz)\| = \|\bM(\by \x \bz)\|$, but $\bx^\dag\by \neq 0$.
\end{enumerate}

The proof in case $1^\circ$ is relatively straightforward. Introduce the orthonormal system $\bu_1 \vc \bx \x \by$, $\bu_2 \vc \bx \x \bz$, $\bu_3 \vc \by \x \bz$ and observe the following chain of (in)equalities:
\begin{align*}
2\| \bE^{p/2}\bu_1 \|^{2/p} & \leq 2\max_{\|\bv\|=1}\| \bE^{p/2}\bv \|^{2/p} = 2\rho(\bE) \leq \Tr(\bE) 
\\
& = \sum_{i=1}^3 \bu_i^\dag \bE \bu_i \leq \sum_{i=1}^3 \left( \bu_i^\dag \bE^{p/2} \bu_i \right)^{2/p} \leq \sum_{i=1}^3 \| \bE^{p/2} \bu_i \|^{2/p},
\end{align*}
where the second inequality is \eqref{war}, the third one follows from Jensen's inequality (by the convexity of the map $t \mapsto t^{p/2}$), and the final one follows from the Cauchy--Schwarz inequality. Subtracting $\| \bE^{p/2}\bu_1 \|^{2/p}$ from both sides yields the triangle inequality \eqref{ineq3}.

In case $2^\circ$, notice first that a vector of unit length orthogonal to both $\bx$ and $\by$ must be of the form 
\begin{align}
\label{zort}
\bz = \ee^{\ii\phi}\frac{\bar{\bx} \x \bar{\by}}{\|\bx \x \by\|}.
\end{align}
Plugging this along with $M_{13} = M_{23} \cv \Lambda$ into the third equation in \eqref{eqs1} yields 
\begin{align*}
\bar{\bx} \x \bM^2((\bar{\bx} \x \bar{\by}) \x \bx) + \bar{\by} \x \bM^2((\bar{\bx} \x \bar{\by}) \x \by) = 2 \Lambda^2(\bar{\bx} \x \bar{\by}).
\end{align*}
By \eqref{id3}, this simplifies to
\begin{align}
\label{eqs3}
\bar{\bx} \x \bM^2\bar{\by} - \bar{\by} \x \bM^2\bar{\bx} - (\by^\dag\bx)\bar{\bx} \x \bM^2\bar{\bx} + (\bx^\dag\by)\bar{\by} \x \bM^2\bar{\by} = 2 \Lambda^2(\bar{\bx} \x \bar{\by}).
\end{align}
Taking inner products with $\bx$, $\by$ on both sides of \eqref{eqs3} yields, with the help of \eqref{id1},
\begin{align*}
& \left\{ \begin{array}{l}
- (\bx \x \by)^\dag \bM^2\bar{\bx} + (\bx^\dag\by)(\bx \x \by)^\dag \bM^2\bar{\by} = 0
\\[8pt]
- (\bx \x \by)^\dag \bM^2\bar{\by} + (\by^\dag\bx)(\bx \x \by)^\dag \bM^2\bar{\bx} = 0
\end{array} \right.
\end{align*}
what can be nicely expressed in matrix form as
\begin{align}
\label{eqs4}
& \begin{bmatrix} -1 & \bx^\dag\by \\ \by^\dag\bx & -1 \end{bmatrix} \begin{bmatrix} (\bx \x \by)^\dag \bM^2\bar{\bx} \\ (\bx \x \by)^\dag \bM^2\bar{\by} \end{bmatrix} = \begin{bmatrix} 0 \\ 0 \end{bmatrix}.
\end{align}
But since $1 - |\bx^\dag\by|^2 = \|\bx \x \by \|^2 \neq 0$ by assumption, equation \eqref{eqs4} yields $(\bx \x \by)^\dag \bM^2\bar{\bx} = (\bx \x \by)^\dag \bM^2\bar{\by} = 0$. In other words, the vector $\bM^2(\bar{\bx} \x \bar{\by})$ is orthogonal to both $\bx$ and $\by$, and hence must be proportional to $\bar{\bx} \x \bar{\by}$, i.e., it must be an \emph{eigenvector} of $\bM^2 = \bE^p$, corresponding to, say, the eigenvalue $\lambda_3^p$ (without loss of generality).

Let $\bv_1$, $\bv_2$ be the other two eigenvectors of $\bE^p$ with eigenvalues $\lambda_1^p$, $\lambda_2^p$, respectively. Bearing in mind \eqref{zort}, the vectors $\bx \x \bz$ and $\by \x \bz$ both lie in $\sspan\{\bv_1,\bv_2\}$ and their norm is 1 (recall we have $\bx^\dag\bz = \by^\dag\bz = 0$ in the considered case), and so we can decompose them as
\begin{align}
\label{eqs5}
& \bx \x \bz = \bv_1 \ee^{\ii \xi_1} \cos\varphi + \bv_2 \ee^{\ii \xi_2} \sin\varphi
\\
\nonumber
& \by \x \bz = \bv_1 \ee^{\ii \eta_1} \cos\theta + \bv_2 \ee^{\ii \eta_2} \sin\theta
\end{align}
for some angles $\xi_1,\xi_2,\varphi,\eta_1,\eta_2,\theta$. With all that in mind, our desired triangle inequality \eqref{ineq3} reduces to
\begin{align*}
\lambda_3 \| \bx \x \by \|^{2/p} \leq \left(\lambda_1^p\cos^2\varphi + \lambda_2^p \sin^2\varphi\right)^{1/p} + \left(\lambda_1^p\cos^2\theta + \lambda_2^p \sin^2\theta\right)^{1/p}.
\end{align*}
The troublesome $\| \bx \x \by \|$ lingering on the left-hand side can be expressed in the language of the above angles, too. Indeed, by \eqref{zort} and \eqref{id4} we have that
\begin{align*}
(\bx \x \bz) \x (\by \x \bz) = \ee^{2\ii\phi} \, \bar{\bx} \x \bar{\by},
\end{align*}
and so, introducing an auxiliary angle $\omega \vc \tfrac{1}{2}(\xi_2-\xi_1+\eta_1-\eta_2)$, after some trigonometry we obtain
\begin{align*}
\| \bx \x \by \|^2 & = \|(\bx \x \bz) \x (\by \x \bz)\|^2 = | \cos\varphi \sin\theta - \ee^{2\ii \omega} \sin\varphi \cos\theta|^2
\\
& = \cos^2\omega \sin^2(\theta-\varphi) + \sin^2\omega \sin^2(\theta+\varphi).
\end{align*}
Notice that the right-hand side can be estimated from above by $\max \{\sin^2(\theta\pm\varphi)\}$, so in order to prove the triangle inequality \eqref{ineq3} it suffices to show that
\begin{align}
\label{eqs6}
(\lambda_1 + \lambda_2) |\sin(\theta\pm\varphi)|^{2/p} \leq \left(\lambda_1^p\cos^2\varphi + \lambda_2^p \sin^2\varphi\right)^{1/p} + \left(\lambda_1^p\cos^2\theta + \lambda_2^p \sin^2\theta\right)^{1/p},
\end{align}
where we have also replaced $\lambda_3$ with $\lambda_1 + \lambda_2$, which by \eqref{war} is the former's upper bound.

Luckily, we do not have to attack \eqref{eqs6} in full generality, as we still have not taken full advantage of the assumption $\|\bM(\bx \x \bz)\| = \|\bM(\by \x \bz)\|$. Indeed, when we substitute \eqref{eqs5} and $\bM = \bE^{p/2}$ into this assumption, after some trigonometry it reduces to
\begin{align}
\label{eqs7}
(\lambda_2^p - \lambda_1^p)\sin(\theta-\varphi)\sin(\theta+\varphi) = 0
\end{align}
and we are thus left with three possibilities.
\begin{itemize}
\item If $\lambda_1 = \lambda_2 \cv \lambda$, then the LHS of \eqref{eqs6} becomes $2\lambda|\sin(\theta\pm\varphi)|^{2/p}$, whereas the RHS reduces to $2\lambda$, and so the inequality clearly holds.
\item If $\sin(\theta-\varphi) = 0$, then $\varphi = \theta + k\pi$ and \eqref{eqs6} boils down to showing that
\begin{align}
\label{eqs8}
(\lambda_1 + \lambda_2)^p \sin^2 2\theta \leq 2^p \left(\lambda_1^p\cos^2\theta + \lambda_2^p \sin^2\theta\right).
\end{align}
Indeed, by the convexity of the map $t \mapsto t^{p/2}$, one can write that 
\begin{align*}
& \left(\tfrac{\lambda_1 + \lambda_2}{2}\right)^p \sin^2 2\theta \leq  \Big(\tfrac{\lambda_1^{p/2}}{2} + \tfrac{\lambda_2^{p/2}}{2}\Big)^2 \sin^2 2\theta 
\\
& = \big(\lambda_1^p\cos^2\theta + \lambda_2^p \sin^2\theta\big) - \big(\lambda_1^{p/2}\cos^2 \theta - \lambda_2^{p/2}\sin^2 \theta \big)^2 \leq \big(\lambda_1^p\cos^2\theta + \lambda_2^p \sin^2\theta\big).
\end{align*}
\item If $\sin(\theta+\varphi) = 0$, then $\varphi = -\theta + k\pi$ and \eqref{eqs6} again boils down to \eqref{eqs8}.
\end{itemize}
This finishes the analysis of case $2^\circ$ and concludes the whole proof.
\end{proof}

\medskip

Let now $n \geq 3$ and recall that $d_p(\bx,\by) \vc \|\bE^{p/2}(\bx \wedge \by)\|^{2/p}$, where $\bE \in \mathcal{B}(\Lambda^2\sC^n)$ is a positive-definite operator defined via $\bE(\be_i \wedge \be_j) \vc E_{ij}\be_i \wedge \be_j$ for any $i,j$, where $(E_{ij})$ is some fixed distance matrix. The following partial result demonstrates that in order to establish the triangle inequality \eqref{triangle1}, it suffices to focus on \emph{orthonormal} systems of vectors.
\begin{proposition}
\label{partialresult}
Triangle inequality \eqref{triangle1} holds if and only if
\begin{align}
\label{triangle2}
\|\bE^{p/2}(\bu \wedge \bv)\|^{2/p} \leq \|\bE^{p/2}(\bu \wedge \bw)\|^{2/p} + \|\bE^{p/2}(\bv \wedge \bw)\|^{2/p}
\end{align}
for any \emph{orthonormal} $\{\bu,\bv,\bw\} \subset \sC^n$.
\end{proposition}
\begin{proof}
Assume \eqref{triangle2} holds. Fix norm-one vectors $\bx, \by, \bz \in \sC^n$ and define $V$ to be any 3-dimensional subspace containing them\footnote{If the system $\{\bx, \by, \bz\}$ is linearly independent, then of course $V$ is simply its span.} along with $\Lambda^2 V$ being the corresponding 3-dimensional subspace of $\Lambda^2\sC^n$. Let $h: \Lambda^2 V \times \Lambda^2 V \rightarrow \sC$ be the restriction of the Hermitian form $\Lambda^2\sC^n \times \Lambda^2\sC^n \ni (\bV,\bW) \mapsto \langle \bV | \bE^p \bW \rangle$, and let $\{\bB_1,\bB_2,\bB_3\}$ be its orthonormal eigenbasis, i.e. a basis of $\Lambda^2 V$ in which the matrix of $h$, denoted by $\bH$, is diagonal, $\bH = \diag(\mu_1,\mu_2,\mu_3)$. We claim that $2 \rho(\bH^{1/p}) \leq \Tr \bH^{1/p}$, what on the strength of Proposition \ref{n3case} would finish the proof.

In order to prove the claim we shall first find an orthonormal basis $\{\bbf_1,\bbf_2,\bbf_3\}$ such that
\begin{align}
\label{hodge}
\bB_1 = \bbf_2 \wedge \bbf_3, \qquad \bB_2 = \bbf_3 \wedge \bbf_1, \qquad \bB_3 = \bbf_1 \wedge \bbf_2,
\end{align}
what is possible due to the three-dimensionality of $V$ with the help of the complex Hodge star operator\footnote{For an alternative, matrix-theoretic justification see \cite[Lemma 3.1]{llaproc}.} (cf. \cite{Wells}). Concretely, let $\{\bv_1,\bv_2,\bv_3\}$ be an orthonormal basis of $V$ such that the matrix $(U_{ij})$ changing the basis from $\{\bB_1,\bB_2,\bB_3\}$ to the `new' basis $\{\bv_2 \wedge \bv_3, \bv_3 \wedge \bv_1, \bv_1 \wedge \bv_2 \}$ has determinant 1 (which can always be ensured by adjusting the phase factors). Define the operator $\star: V \rightarrow \Lambda^2 V$ by (linearly extending) the formulas
\begin{align*}
\star\bv_1 \vc \bar{\bv}_2 \w \bar{\bv}_3, \qquad \star\bv_2 \vc \bar{\bv}_3 \w \bar{\bv}_1, \qquad \star\bv_3 \vc \bar{\bv}_1 \w \bar{\bv}_2.
\end{align*}
Clearly, $\star$ is unitary, satisfies
\begin{align}
\label{hodge2}
\langle \overline{\star\bv_i} | \bv_j \w \bv_k \rangle = \varepsilon_{ijk}, \qquad i,j,k = 1,2,3
\end{align}
where $\varepsilon_{ijk}$ is the Levi-Civita symbol, and allows us to express the new basis as $\{\overline{\star\bv_1},\overline{\star\bv_2},\overline{\star\bv_3}\}$, along with the entries of the change-of-basis matrix as $U_{ij} = \langle \overline{\star\bv_j} | \bB_i \rangle$.

Now, set $\bbf_l \vc \star^{-1} \bar{\bB}_l = \sum_m \bar{U}_{lm} \bv_m$, $l=1,2,3$. By the unitarity of $\star$, thus defined system $\{\bbf_1,\bbf_2,\bbf_3\}$ is orthonormal. To see that it satisfies \eqref{hodge}, take, e.g., $\bbf_2 \w \bbf_3$ and calculate its inner products with $\bB_l$, $l=1,2,3$, obtaining
\begin{align*}
& \langle \bB_l | \bbf_2 \w \bbf_3 \rangle = \langle \sumka_i U_{li} \overline{\star\bv_i} | (\sumka_j \bar{U}_{2j} \bv_j) \w (\sumka_k \bar{U}_{3k} \bv_k) \rangle = \sumka_{ijk} \bar{U}_{li}\bar{U}_{2j}\bar{U}_{3k}\varepsilon_{ijk} = \begin{cases} 1 & \textnormal{for } l=1 \\ 0 & \textnormal{otherwise} \end{cases},
\end{align*}
where we have expanded $\bB_l$ in the basis $\{\overline{\star\bv_i}\}$, used \eqref{hodge2} and recognized the formula for the (complex conjugate of the) determinant of a $3 \times 3$ matrix. This proves that indeed $\bbf_2 \w \bbf_3 = \bB_1$ and the other two equalities in \eqref{hodge} can be shown similarly.

Having established \eqref{hodge}, we can now write
\begin{align*}
&\mu_1 = h(\bB_1,\bB_1) = \|\bE^{p/2}(\bbf_2 \wedge \bbf_3)\|^2,
\\
&\mu_2 = h(\bB_2,\bB_2) = \|\bE^{p/2}(\bbf_1 \wedge \bbf_3)\|^2,
\\
&\mu_3 = h(\bB_3,\bB_3) = \|\bE^{p/2}(\bbf_1 \wedge \bbf_2)\|^2,
\end{align*}
and our assumption \eqref{triangle2} immediately yields $2\max_i \mu_i^{1/p} \leq \sum_j \mu_j^{1/p}$, which is exactly our desired condition $2 \rho(\bH^{1/p}) \leq \Tr \bH^{1/p}$.
\end{proof}

\section{An auxiliary convexity result and the proof of the triangle inequality}
\label{sec::3}

The proof of Theorem \ref{main} will be a consequence of the following convexity result involving the squared moduli of certain minors of order 2 and 3.
\begin{theorem}
\label{convexThm}
Fix $n \geq 3$. For any symmetric matrix $(a_{ij}) \in \mathbb{M}_n(\sR)$ with nonnegative entries, any orthonormal system $\{\bx,\by,\bz\} \subset \sC^n$ and any jointly convex, totally symmetric map $f: \mathbb{R}_{\geq 0}^3 \rightarrow \mathbb{R}$, the following inequality holds
\begin{align}
\label{convex1}
f \Big( \sum_{i<j}a_{ij}\left|\left|\begin{smallmatrix}x_i & x_j \\ y_i & y_j \end{smallmatrix}\right|\right|^2, \sum_{i<k}a_{ik}\left|\left|\begin{smallmatrix}x_i & x_k \\ z_i & z_k \end{smallmatrix}\right|\right|^2, \sum_{j<k}a_{jk}\left|\left|\begin{smallmatrix}y_j & y_k \\ z_j & z_k \end{smallmatrix}\right|\right|^2 \Big) \leq \sum_{i<j<k}f(a_{ij},a_{ik},a_{jk}) \left|\left|\begin{smallmatrix}x_i & x_j & x_k \\ y_i & y_j & y_k \\ z_i & z_j & z_k \end{smallmatrix}\right|\right|^2.
\end{align}
\end{theorem}
In order to prove the above theorem, we shall need the following lemma.

\begin{lemma}
\label{minorialThm}
Fix $n \geq 3$. For any symmetric matrix $(a_{ij}) \in \mathbb{M}_n(\sR)$ with nonnegative entries and any orthonormal system $\{\bx,\by,\bz\} \subset \sC^n$,
\begin{align}
\label{minorial}
\sum_{i<j<k}\min\{a_{ij},a_{ik},a_{jk}\}\left|\left| \begin{smallmatrix} x_i & x_j & x_k \\ y_i & y_j & y_k \\ z_i & z_j & z_k \end{smallmatrix} \right|\right|^2 \leq \sum_{i<j}a_{ij}\left|\left| \begin{smallmatrix} x_i & x_j \\ y_i & y_j \end{smallmatrix} \right|\right|^2 \leq \sum_{i<j<k}\max\{a_{ij},a_{ik},a_{jk}\}\left|\left| \begin{smallmatrix} x_i & x_j & x_k \\ y_i & y_j & y_k \\ z_i & z_j & z_k \end{smallmatrix} \right|\right|^2.
\end{align}
\end{lemma}
\begin{proof}
Fix the orthonormal system $\{\bx,\by,\bz\}$ and, without loss of generality, assume that $a_{ii} = 0$ for all $i$. Observe, crucially, that both inequalities in \eqref{minorial} are \emph{piecewise-linear} with respect to $\bba \vc (a_{ij})_{i<j} \in \sR_+^{\binom{n}{2}}$. More specifically, let us partition the nonnegative $\binom{n}{2}$-orthant $\sR_{\geq 0}^{\binom{n}{2}}$ into $\binom{n}{2}!$ regions $C_\sigma$ (where $\sigma$ runs over $S_{\binom{n}{2}}$) defined as
\begin{align*}
C_\sigma \vc \{ \bP_\sigma \bba \in \sR_{\geq 0}^{\binom{n}{2}} \, | \, a_{12} \geq a_{13} \geq \ldots \geq a_{n-1,n} \geq 0 \},
\end{align*}
where $\bP_\sigma$ is the permutation matrix representing $\sigma \in S_{\binom{n}{2}}$. Then on each region $C_\sigma$ inequalities \eqref{minorial} become linear, because for every $i<j<k$ both the minimum on the left-hand side and the maximum on the right-hand side are determined by the region-specific ordering of the components of $\bba$. Notice, moreover, that each region $C_\sigma$ is actually a \emph{polyhedral cone}, and hence it is generated by finitely many vectors \cite[Corollary 1.3.13]{LHK13}. Therefore, proving that a linear inequality holds on $C_\sigma$ amounts to veryfing it on its generators, which can be chosen to be the following $\binom{n}{2}$ zero-one vectors
\begin{align*}
\bP_\sigma [\underbrace{1,\ldots,1}_k,\underbrace{0,\ldots,0}_{\binom{n}{2} - k}]^\top, \quad k=1,\ldots,\binom{n}{2}.
\end{align*}
Bearing in mind that $\sigma$ runs over all permutations, the above observations can be summarized as follows: \emph{In order to prove \eqref{minorial}, $(a_{ij})$ can be assumed to be a zero-one matrix}. This means that, if we introduce $\cS \vc \{ \{i,j\} \, | \, a_{ij} = 1 \}$, it suffices to prove
\begin{align}
\label{minorial1}
\sum_{\substack{i<j<k \\ \{i,j\},\{i,k\},\{j,k\} \in \cS}}\left|\left| \begin{smallmatrix} x_i & x_j & x_k \\ y_i & y_j & y_k \\ z_i & z_j & z_k \end{smallmatrix} \right|\right|^2 \leq \sum_{\substack{i<j \\ \{i,j\} \in \cS}}\left|\left| \begin{smallmatrix} x_i & x_j \\ y_i & y_j \end{smallmatrix} \right|\right|^2 \leq \sum_{\substack{i<j<k \\ \{i,j\} \in \cS \, \vee \, \{i,k\} \in \cS \, \vee \, \{j,k\} \in \cS}}\left|\left| \begin{smallmatrix} x_i & x_j & x_k \\ y_i & y_j & y_k \\ z_i & z_j & z_k \end{smallmatrix} \right|\right|^2.
\end{align}

Let us focus on the first inequality. It is worthwhile to express it in the language of exterior algebra\footnote{The wedge product of $k$ vectors is defined here as $\bv_1 \wedge \ldots \wedge \bv_k \vc \frac{1}{\sqrt{k!}} \sum_{\sigma \in S_k} \sgn\sigma \ \bv_{\sigma_1} \otimes \ldots \otimes \bv_{\sigma_k}$.}. To this end, define the projection operators $P \in \mathcal{B}(\Lambda^2\mathbb{C}^n)$, $Q \in \mathcal{B}(\Lambda^3\mathbb{C}^n)$ via
\begin{align*}
P \vc \sum_{i<j} \chi_{ij} |\be_i \wedge \be_j \rangle \langle \be_i \wedge \be_j |, \qquad Q \vc \sum_{i<j<k} \chi_{ij}\chi_{ik}\chi_{jk}|\be_i \wedge \be_j \wedge \be_k \rangle \langle \be_i \wedge \be_j \wedge \be_k |,
\end{align*}
where $\chi_{ij} = 1$ for $\{i,j\} \in \cS$ and $0$ otherwise. The first inequality in \eqref{minorial1} is now nothing but
\begin{align}
\label{minorial2}
\|Q(\bx \wedge \by \wedge \bz)\|^2 \leq \|P(\bx \wedge \by)\|^2,
\end{align}
and we shall in fact prove a stronger result, namely,
\begin{align}
\label{minorial3}
\forall \bv \in \sC^n \ \forall \bB \in \Lambda^2\mathbb{C}^n \quad \|Q(\bB \wedge \bv)\|^2 \leq \|P\bB\|^2\|\bv\|^2.
\end{align}
Indeed, one has that
\begin{align*}
\bB \wedge \bv & = \Big(\sum_{i<j} B_{ij} \be_i \wedge \be_j \Big) \wedge \Big(\sum_k v_{k} \be_k \Big) = \tfrac{1}{2} \sum_{i,j,k} B_{ij}v_k \, \be_i \wedge \be_j \wedge \be_k
\\
& = \sum_{i<j<k} (B_{ij}v_k - B_{ik}v_j + B_{jk}v_i) \, \be_i \wedge \be_j \wedge \be_k,
\end{align*}
and hence
\begin{align}
\label{minorialQ}
\|Q(\bB \wedge \bv)\|^2 = \sum_{i<j<k} \chi_{ij}\chi_{ik}\chi_{jk} |B_{ij}v_k - B_{ik}v_j + B_{jk}v_i|^2.
\end{align}
On the other hand, one has that
\begin{align*}
P(\bB) \wedge \bv & = \Big(\sum_{i<j} \chi_{ij}B_{ij} \be_i \wedge \be_j \Big) \wedge \Big(\sum_k v_{k} \be_k \Big) = \tfrac{1}{2} \sum_{i,j,k} \chi_{ij}B_{ij}v_k \, \be_i \wedge \be_j \wedge \be_k
\\
& = \sum_{i<j<k} (\chi_{ij}B_{ij}v_k - \chi_{ik}B_{ik}v_j + \chi_{jk}B_{jk}v_i) \, \be_i \wedge \be_j \wedge \be_k,
\end{align*}
and hence
\begin{align}
\label{minorialP}
\|P(\bB) \wedge \bv\|^2 = \sum_{i<j<k} |\chi_{ij}B_{ij}v_k - \chi_{ik}B_{ik}v_j + \chi_{jk}B_{jk}v_i|^2.
\end{align}
But since $\chi_{ij}\chi_{ik}\chi_{jk} |B_{ij}v_k - B_{ik}v_j + B_{jk}v_i|^2 \leq |\chi_{ij}B_{ij}v_k - \chi_{ik}B_{ik}v_j + \chi_{jk}B_{jk}v_i|^2$ for any $i,j,k$, we obtain that $\|Q(\bB \wedge \bv)\|^2 \leq \|P(\bB) \wedge \bv\|^2$, which in turn can be bounded from above by $\|P\bB\|^2\|\bv\|^2$, for instance on the strength of the identity\footnote{This generalization of the Lagrange identity can be easily proven by direct calculation in an orthonormal basis of $\mathbb{C}^n$ containing $\bw/\|\bw\|$ as its first element.}
\begin{align*}
\forall \bw \in \mathbb{C}^n \ \forall \bOmega \in \Lambda^k\mathbb{C}^n \quad  \|\bw \wedge \bOmega \|^2 = \|\bw\|^2 \|\bOmega\|^2 - \|\bw \lrcorner \, \bOmega\|^2
\end{align*}
for any $k \in \{1,\ldots,n\}$, where $\lrcorner$ denotes the interior product, defined on simple $k$-vectors via $\bw \lrcorner \, (\bv_1\wedge \ldots \wedge \bv_k) = \sum_i (-1)^{i-1} \langle \bw | \bv_i \rangle \bv_1 \wedge \ldots \wedge \widehat{\bv}_i \wedge \ldots \wedge \bv_k$. This completes the proof of \eqref{minorial3} and thus of the first inequality in \eqref{minorial1}.

As for the second inequality, observe that, by the Cauchy--Binet theorem and by the orthonormality of $\{\bx,\by,\bz\}$, it can be equivalently written as
\begin{align*}
1 - \sum_{\substack{i<j \\ \{i,j\} \in \cS^c}}\left|\left| \begin{smallmatrix} x_i & x_j \\ y_i & y_j \end{smallmatrix} \right|\right|^2 \leq 1 - \sum_{\substack{i<j<k \\ \{i,j\}, \{i,k\}, \{j,k\} \in \cS^c}}\left|\left| \begin{smallmatrix} x_i & x_j & x_k \\ y_i & y_j & y_k \\ z_i & z_j & z_k \end{smallmatrix} \right|\right|^2
\end{align*}
and since the subset of pairs $\cS$ was arbitrary, the above inequality follows from the one already proven. This completes the proof on the entire lemma.
\end{proof}

\begin{proof}[Proof of Theorem \ref{convexThm}]
Let us first introduce some convenient notation. Define 
\begin{align*}
\bg \vc \Big( \sum_{i<j}a_{ij}\left|\left|\begin{smallmatrix}x_i & x_j \\ y_i & y_j \end{smallmatrix}\right|\right|^2, \sum_{i<k}a_{ik}\left|\left|\begin{smallmatrix}x_i & x_k \\ z_i & z_k \end{smallmatrix}\right|\right|^2, \sum_{j<k}a_{jk}\left|\left|\begin{smallmatrix}y_j & y_k \\ z_j & z_k \end{smallmatrix}\right|\right|^2 \Big) \in \mathbb{R}^3_{\geq 0},
\end{align*}
so that the left-hand side of our desired inequality \eqref{convex1} becomes simply $f(\bg)$.

Let us introduce the set of ordered triples $\cO_3 \vc \{ (i,j,k) \ | \ 1 \leq i < j < k \leq n \}$ and, for any $t \vc (i,j,k) \in \cO_3$, let us denote $p_t \vc \left|\left| \begin{smallmatrix} x_i & x_j & x_k \\ y_i & y_j & y_k \\ z_i & z_j & z_k \end{smallmatrix} \right|\right|^2$. Notice that this defines a probability distribution on $\cO_3$, because $\sum_t p_t = 1$ by the orthonormality of $\{\bx,\by,\bz\}$.

For any ordered triple $t \vc (i,j,k)$ define $\bba_t \vc (a_{ij},a_{ik},a_{jk}) \in \sR^3_{\geq 0}$ and observe that the right-hand side of our desired inequality \eqref{convex1} is nothing but $\sum_t p_t f(\bba_t)$.

The crucial step is to prove that it is always possible to find a family of $3 \times 3$ doubly stochastic matrices $\{\bQ_t\}_{t \in \cO_3}$ such that
\begin{align}
\label{convex2}
\sum_t p_t \bQ_t \bba_t = \bg.
\end{align}
Once this claim is proven, inequality \eqref{convex1} follows. Indeed, since by the Birkhoff--von Neumann theorem every $\bQ_t$ is a convex combination of permutation matrices $\bP_\sigma$, we would have
\begin{align*}
f(\bg) & = f\Big(\sum\nolimits_t p_t \bQ_t \bba_t\Big) = f\Big(\sum\nolimits_t p_t \sum\nolimits_{\sigma \in S_3} \lambda_{t,\sigma} \bP_\sigma \bba_t\Big) 
\\
& \leq \sum\nolimits_t p_t \sum\nolimits_{\sigma \in S_3} \lambda_{t,\sigma} f\big( \bP_\sigma \bba_t\big) = \sum\nolimits_t p_t \sum\nolimits_{\sigma \in S_3} \lambda_{t,\sigma} f(\bba_t) = \sum\nolimits_t p_t f(\bba_t),
\end{align*}
where we have used \eqref{convex2}, Birkhoff--von Neumann theorem, Jensen's inequality by the joint convexity of $f$, the total symmetricity of $f$ and, in the final equality, we simply summed over $\sigma$ the convex combination coefficients $\lambda_{t,\sigma}$.

In order to prove the claim, observe that \eqref{convex2} can be restated as saying that $\bg$ belongs to the following Minkowski combination of compact convex sets
\begin{align*} 
\bg \in \K \vc \sum_t p_t \{ \bQ \bba_t \, | \, \bQ \in \B_3 \},
\end{align*}
where $\B_3$ denotes the Birkhoff polytope of $3$-by-$3$ doubly stochastic matrices. But since $\K$ is thus compact and convex itself, the condition $\bg \in \K$ in turn can be equivalently expressed via its support function\footnote{See e.g. \cite{Schneider} for the exposition of the basic properties of support functions.} $h_\K(\bw) \vc \sup_{\bu \in \K} \bw^\top \bu$ as
\begin{align*}
\forall \bw \in \sR^3 \quad \bw^\top\bg \leq h_\K(\bw).
\end{align*}
Since support functions of compact convex sets satisfy $h_{\lambda\C} = \lambda h_\C$ for any $\lambda \geq 0$ and are Minkowski additive \cite[Theorem 1.7.5]{Schneider}, the function $h_\K$ can be expressed as
\begin{align*}
h_\K(\bw) = \sum_t p_t \sup_{Q \in \B_3} \bw^\top\bQ \bba_t = \sum_t p_t \max_{\sigma \in S_3} \, \bw^\top\bP_\sigma \bba_t = \sum_t p_t (\bw^\downarrow)^\top\bba^\downarrow_t,
\end{align*}
where in the second equality we have used the fact that the supremum of the linear map $\B_3 \ni \bQ \mapsto \bw^\top\bQ \bba_t$ must be attained at one of the extreme points of $\B_3$---which by the Birkhoff--von Neumann theorem are precisely the permutation matrices---and the third equality simply picks the rearrangement maximum for every $t \in \cO_3$.

But since $\bw^\top\bg \leq (\bw^\downarrow)^\top\bg^\downarrow$ by the rearrangement inequality, we can restrict our attention to the set $C \vc \{ \bw \in \sR^3 \, | \, w_1 \geq w_2 \geq w_3 \}$ and prove the \emph{linear} inequality
\begin{align}
\label{convex3}
\forall \bw \in C \quad \bw^\top \bg^\downarrow \leq \sum_t p_t \bw^\top \bba^\downarrow_t.
\end{align}
In fact, since $C$ is a finitely generated cone (cf. the proof of Lemma \ref{minorialThm} above), it suffices to prove \eqref{convex3} on its generators, which can be chosen to be $\bw_1 \vc [1,1,1]^\top, \bw_2 \vc [1,0,0]^\top$ and $\bw_3 \vc [0,0,-1]^\top$.

For $\bw = \bw_1$, \eqref{convex3} is in fact an \emph{equality}. Indeed, by relabeling the dummy indices, moving between the ordered and unordered sums, and making use of the symmetries of $(a_{ij})$ and of the minors, one can write that
\begin{align*}
\bw_1^\top \bg^\downarrow & = \sum_{i<j}a_{ij}\left|\left|\begin{smallmatrix}x_i & x_j \\ y_i & y_j \end{smallmatrix}\right|\right|^2 + \sum_{i<k}a_{ik}\left|\left|\begin{smallmatrix}x_i & x_k \\ z_i & z_k \end{smallmatrix}\right|\right|^2 + \sum_{j<k}a_{jk}\left|\left|\begin{smallmatrix}y_j & y_k \\ z_j & z_k \end{smallmatrix}\right|\right|^2
\\
& = \tfrac{1}{2} \sum_{i,j}a_{ij}\left( \left|\left|\begin{smallmatrix}x_i & x_j \\ y_i & y_j \end{smallmatrix}\right|\right|^2 + \left|\left|\begin{smallmatrix}x_i & x_j \\ z_i & z_j \end{smallmatrix}\right|\right|^2 + \left|\left|\begin{smallmatrix}y_i & y_j \\ z_i & z_j \end{smallmatrix}\right|\right|^2 \right) = \tfrac{1}{2} \sum_{i,j,k}a_{ij}\left|\left| \begin{smallmatrix} x_i & x_j & x_k \\ y_i & y_j & y_k \\ z_i & z_j & z_k \end{smallmatrix} \right|\right|^2 
\\
& = \tfrac{1}{6} \sum_{i,j,k}\left(a_{ij} + a_{ik} + a_{jk}\right)\left|\left| \begin{smallmatrix} x_i & x_j & x_k \\ y_i & y_j & y_k \\ z_i & z_j & z_k \end{smallmatrix} \right|\right|^2 = \sum_{i<j<k}\left(a_{ij} + a_{ik} + a_{jk}\right) \left|\left| \begin{smallmatrix} x_i & x_j & x_k \\ y_i & y_j & y_k \\ z_i & z_j & z_k \end{smallmatrix} \right|\right|^2 = \sum_t p_t \bw_1^\top\bba^\downarrow_t.
\end{align*}

For $\bw = \bw_2$ \eqref{convex3} can be written as
\begin{align*}
\max\left\{ \sum_{i<j}a_{ij}\left|\left|\begin{smallmatrix}x_i & x_j \\ y_i & y_j \end{smallmatrix}\right|\right|^2, \sum_{i<k}a_{ik}\left|\left|\begin{smallmatrix}x_i & x_k \\ z_i & z_k \end{smallmatrix}\right|\right|^2, \sum_{j<k}a_{jk}\left|\left|\begin{smallmatrix}y_j & y_k \\ z_j & z_k \end{smallmatrix}\right|\right|^2 \right\} \leq \sum_{i<j<k}\max\{a_{ij},a_{ik},a_{jk}\}\left|\left| \begin{smallmatrix} x_i & x_j & x_k \\ y_i & y_j & y_k \\ z_i & z_j & z_k \end{smallmatrix} \right|\right|^2,
\end{align*}
what follows immediately from the second inequality in \eqref{minorial}.

Similarly, for $\bw = \bw_3$, \eqref{convex3} boils down to
\begin{align*}
\min\left\{ \sum_{i<j}a_{ij}\left|\left|\begin{smallmatrix}x_i & x_j \\ y_i & y_j \end{smallmatrix}\right|\right|^2, \sum_{i<k}a_{ik}\left|\left|\begin{smallmatrix}x_i & x_k \\ z_i & z_k \end{smallmatrix}\right|\right|^2, \sum_{j<k}a_{jk}\left|\left|\begin{smallmatrix}y_j & y_k \\ z_j & z_k \end{smallmatrix}\right|\right|^2 \right\} \geq \sum_{i<j<k}\min\{a_{ij},a_{ik},a_{jk}\}\left|\left| \begin{smallmatrix} x_i & x_j & x_k \\ y_i & y_j & y_k \\ z_i & z_j & z_k \end{smallmatrix} \right|\right|^2,
\end{align*}
what follows immediately from the first inequality in \eqref{minorial}. Invoking now Lemma \ref{minorialThm} establishes that $\bg \in \K$ and thus completes the proof of the theorem. 
\end{proof}

By replacing $f$ with $-f$ in Theorem \ref{convexThm}, we immediately obtain the related concavity result.
\begin{corollary}
\label{concaveCor}
Fix $n \geq 3$. For any symmetric matrix $(a_{ij}) \in \mathbb{M}_n(\sR)$ with nonnegative entries, any orthonormal system $\{\bx,\by,\bz\} \subset \sC^n$ and any jointly concave, totally symmetric map $f: \mathbb{R}_{\geq 0}^3 \rightarrow \mathbb{R}$, the following inequality holds
\begin{align}
\label{concave1}
f \Big( \sum_{i<j}a_{ij}\left|\left|\begin{smallmatrix}x_i & x_j \\ y_i & y_j \end{smallmatrix}\right|\right|^2, \sum_{i<k}a_{ik}\left|\left|\begin{smallmatrix}x_i & x_k \\ z_i & z_k \end{smallmatrix}\right|\right|^2, \sum_{j<k}a_{jk}\left|\left|\begin{smallmatrix}y_j & y_k \\ z_j & z_k \end{smallmatrix}\right|\right|^2 \Big) \geq \sum_{i<j<k}f(a_{ij},a_{ik},a_{jk}) \left|\left|\begin{smallmatrix}x_i & x_j & x_k \\ y_i & y_j & y_k \\ z_i & z_j & z_k \end{smallmatrix}\right|\right|^2.
\end{align}
\end{corollary}

We are now finally ready to prove the triangle inequality for $d_p$.

\begin{proof}[Proof of Theorem \ref{main}]
On the strength of Remark \ref{rem3} and Proposition \ref{partialresult}, it suffices to show \eqref{triangle1} for any fixed orthonormal system $\{\bx,\by,\bz\}$, which can be equivalently stated as
\begin{align}
\label{triangle3}
2\max \{d_p(\bx,\by), d_p(\bx,\bz), d_p(\by,\bz)\} \leq d_p(\bx,\by) + d_p(\bx,\bz) + d_p(\by,\bz)
\end{align}
for any orthonormal $\{\bx,\by,\bz\}$. To this end, let us write the following chain of (in)equalities
\begin{align*}
& 2^p\max \{d^p_p(\bx,\by), d^p_p(\bx,\bz), d^p_p(\by,\bz)\}
\\
& = 2^p\max\left\{ \sum_{i<j}E^p_{ij}\left|\left|\begin{smallmatrix}x_i & x_j \\ y_i & y_j \end{smallmatrix}\right|\right|^2, \sum_{i<k}E^p_{ik}\left|\left|\begin{smallmatrix}x_i & x_k \\ z_i & z_k \end{smallmatrix}\right|\right|^2, \sum_{j<k}E^p_{jk}\left|\left|\begin{smallmatrix}y_j & y_k \\ z_j & z_k \end{smallmatrix}\right|\right|^2 \right\} 
\\
& \leq 2^p\sum_{i<j<k}\max(E^p_{ij},E^p_{ik},E^p_{jk})\left|\left|\begin{smallmatrix}x_i & x_j & x_k \\ y_i & y_j & y_k \\ z_i & z_j & z_k \end{smallmatrix}\right|\right|^2 \leq \sum_{i<j<k}(E_{ij} + E_{ik} + E_{jk})^p \left|\left|\begin{smallmatrix}x_i & x_j & x_k \\ y_i & y_j & y_k \\ z_i & z_j & z_k \end{smallmatrix}\right|\right|^2
\\
& \leq \left( \Big( \sum_{i<j}E^p_{ij}\left|\left|\begin{smallmatrix}x_i & x_j \\ y_i & y_j \end{smallmatrix}\right|\right|^2 \Big)^{1/p} + \Big( \sum_{i<k}E^p_{ik}\left|\left|\begin{smallmatrix}x_i & x_k \\ z_i & z_k \end{smallmatrix}\right|\right|^2 \Big)^{1/p} + \Big( \sum_{j<k}E^p_{jk}\left|\left|\begin{smallmatrix}y_j & y_k \\ z_j & z_k \end{smallmatrix}\right|\right|^2 \Big)^{1/p} \right)^p
\\
& = \left( d_p(\bx,\by) + d_p(\bx,\bz) + d_p(\by,\bz) \right)^p,
\end{align*}
where the first inequality follows from Theorem \ref{convexThm} applied to the convex map $f(u_1,u_2,u_3) \vc \max\{u_1,u_2,u_3\}$ and $a_{ij} \vc E_{ij}^p$ (or in fact directly from Lemma \ref{minorialThm}), the second inequality follows from $(E_{ij})$ being a distance matrix, and the third inequality follows from Corollary \ref{concaveCor} applied to the concave map $f(u_1,u_2,u_3) \vc \big(u_1^{1/p}+u_2^{1/p}+u_3^{1/p}\big)^{p}$ and $a_{ij} \vc E_{ij}^p$.

With \eqref{triangle3} proven, the proof of Theorem \ref{main} is complete and thus $d_p$ given by \eqref{dp} is indeed a distance function on $\PCn$ for any $p \geq 2$.
\end{proof}

\section{Extending the distances $d_p$ to mixed states}
\label{sec::4}

As mentioned in the introduction, the distances $d_p$ first arose as pure-state restrictions of quantum $p$-Wasserstein semidistances proposed in \cite{FECZ21}, cf. formula \eqref{wasserstein_pure} above. However, leaving aside this original optimal transport setting, one might ask whether and how $d_p$ could be extended to a bona fide distance on the entire $\Omega_n$ for any fixed $p \geq 2$ and an underlying distance matrix $(E_{ij})$. In other words, is there a concise way of defining a distance function $D$ on $\Omega_n$ (inducing the standard topology) so as to have $D(\bx \bx^\dag,\by \by^\dag) = d_p(\bx,\by)$ for all $\bx,\by \in \PCn$?

In general, such questions turn out to be highly nontrivial and remain an active area of research in quantum information, as illustrated, e.g., by the recent work by Beatty and Stilck Fran\c{c}a \cite{BeattyFranca} and of Borsoni \cite{Borsoni}. The former defines order-$p$ quantum Wasserstein (semi)distances\footnote{Different from $W_p$ given by \eqref{wasserstein}.} through a suitably adapted convex roof construction\footnote{A special case of this approach was considered earlier by T\'oth and Pitrik \cite{TothPitrik23}.}. The latter article employs a broader framework based on Choquet's theory, introducing the folded Kantorovich semidistance along with the folded Wasserstein pseudodistance and then applying them in the quantum setting. However, the Beatty--Fran\c{c}a semidistances---equivalently, Borsoni's quantum folded Kantorovich semidistances---are not known to satisfy the triangle inequality in full generality. In \cite{BeattyFranca,Borsoni} only certain sufficient conditions have been given, while in \cite{MillerSep} the triangle inequality has been established for the $p=2$ Beatty--Fran\c{c}a semidistance constructed upon the pure-state distance $d_{\text{HS}}$.

With that in mind, below we shall restrict to the following, more specific question (which in light of \cite[Theorem 3.5]{Borsoni} is relevant also to the two abovementioned approaches): Can $d_p$ be extended to a true distance on $\Omega_n$ of the form $(\rho,\sigma) \mapsto \| \rho - \sigma \|^\nu_{H_n^0}$ with $\nu > 0$ and $\|.\|_{H_n^0}$ being \emph{some} Hilbertian norm on the $\sR$-linear space $H_n^0$ of $n \times n$ traceless Hermitian matrices?

The following result shows that it is indeed possible for a certain class of distance matrices $(E_{ij})$, although it singles out a concrete form of $\|.\|_{H_n^0}$ as well as the specific value of $\nu = 2/p$.

\begin{theorem}
\label{mixedthm}
Let $\|.\|_{H_n^0}$ be a Hilbertian norm on $H_n^0$ and let $D_\nu$ be a distance function on $\Omega_n$ given by $D_\nu(\rho, \sigma) = \| \rho - \sigma \|^\nu_{H_n^0}$ for any density matrices $\rho,\sigma$. Then the map $\bx \mapsto \bx \bx^\dag$ is an isometric embedding of the metric space $(\PCn, d_p)$ into $(\Omega_n,D_\nu)$ iff $\nu = 2/p$, $E_{ij} = \|\bv_i-\bv_j\|^{2/p}$ for some affinely independent $\{\bv_1,\ldots,\bv_n\} \subset \sR^m$, and the norm $\|.\|_{H_n^0}$ is given by, for any $\bDelta \in H_n^0$,
\begin{align}
\label{H0norm}
\|\bDelta\|_{H_n^0} = \sqrt{ \sum\limits_{i<j} E^p_{ij} \left( |\Delta_{ij}|^2 - \Delta_{ii}\Delta_{jj} \right)} = \sqrt{ \sum\limits_{i<j} \|\bv_i-\bv_j\|^2 \left( |\Delta_{ij}|^2 - \Delta_{ii}\Delta_{jj} \right)}.
\end{align}
\end{theorem}

In order to prove the above theorem, we shall need the following lemma.
\begin{lemma}
\label{mixedlem}
Let $(A_{ij})$ be an $n \times n$ real symmetric hollow matrix. The following conditions are equivalent
\begin{enumerate}[(i)]
\item $A_{ij} =  \|\bv_i-\bv_j\|^2$ for some $\{\bv_1,\ldots,\bv_n\} \subset \sR^m$.
\item $\sum_{i<j} A_{ij} \left(|\Delta_{ij}|^2-\Delta_{ii}\Delta_{jj}\right) \geq 0$ for all $\bDelta \in H^0_n$.
\end{enumerate}
\end{lemma}
\begin{proof}
In order to show that (i) implies (ii), take any $\bDelta \in H^0_n$ and observe that
\begin{align*}
\sum_{i<j} \|\bv_i-\bv_j\|^2 \Delta_{ii}\Delta_{jj} = \underbrace{\sum_{i,j} \|\bv_i\|^2 \Delta_{ii}\Delta_{jj}}_{=\, 0} - \sum_{i,j} ( \bv_i \cdot \bv_j ) \Delta_{ii}\Delta_{jj} = - \left\| \sum\nolimits_k \Delta_{kk} \bv_k \right\|^2,
\end{align*}
and so the sum in (ii) can rewritten as
\begin{align}
\label{mixedlem1}
\sum_{i<j} \|\bv_i-\bv_j\|^2 \left(|\Delta_{ij}|^2-\Delta_{ii}\Delta_{jj}\right) = \sum_{i<j} \|\bv_i-\bv_j\|^2 |\Delta_{ij}|^2 + \left\| \sum\nolimits_k \Delta_{kk} \bv_k \right\|^2,
\end{align}
which is manifestly nonnegative.

To show the converse implication, suppose that (i) is \emph{not} satisfied. Then the celebrated Schoenberg criterion \cite{Schoenberg} assures there exists a vector $\bw \in \sR^n$ such that $\sum_k w_k = 0$ and $\sum_{i,j} A_{ij}w_iw_j > 0$. But then condition (ii) is violated by $\bDelta = \diag(w_1,\ldots,w_n)$, because in that case
\begin{align*}
\sum_{i<j} A_{ij} \left(|\Delta_{ij}|^2-\Delta_{ii}\Delta_{jj}\right) = -\sum_{i<j} A_{ij} w_i w_j < 0.
\end{align*}
\end{proof}

\begin{proof}[Proof of Theorem \ref{mixedthm}]
The assumption that $\bx \mapsto \bx \bx^\dag$ is an isometric embedding means that
\begin{align}
\label{mixed1}
\| \bx \bx^\dag - \by \by^\dag \|^\nu_{H_n^0} = d_p(\bx,\by)
\end{align}
for all $\bx,\by \in \PCn$. In particular, for the family of pure states $\bz(t) := \cos t \be_1 + \sin t \be_2$ parametrized by $t > 0$ one obtains, by substituting $\bx = \bz(t)$ and $\by = \bz(0)$ and dividing both sides by $t^\nu$,
\begin{align*}
\frac{\|\sin^2 t (\be_2\be_2^\dag - \be_1\be_1^\dag) + \sin t \cos t (\be_1\be_2^\dag + \be_2\be_1^\dag)\|^\nu_{H_n^0}}{t^\nu} = E_{12} \frac{\sin^{2/p} t}{t^\nu}.
\end{align*}
Notice now that the left-hand side has a finite positive limit as $t \rightarrow 0^+$ (equal to $\|\be_1\be_2^\dag + \be_2\be_1^\dag\|^\nu_{H_n^0}$), and that already forces the equality $\nu = 2/p$ in order to obtain a finite positive limit on the right-hand side as well (equal to $E_{12}$).

With the above in mind, identity \eqref{mixed1} can be now rewritten as
\begin{align}
\label{mixed2}
\| \bx \bx^\dag - \by \by^\dag \|^2_{H_n^0} = d_p^p(\bx,\by) = \sum\limits_{i<j} E^p_{ij} |x_iy_j - x_jy_i|^2.
\end{align}
It is now crucial to observe that each expression $|x_iy_j - x_jy_i|^2$ can be rewritten in terms of the entries of the traceless matrix $\bx \bx^\dag - \by \by^\dag$. Concretely,
\begin{align}
\label{mixed3}
|x_iy_j - x_jy_i|^2 = |x_i \bar{x_j} - y_i \bar{y_j}|^2 - (|x_i|^2 - |y_i|^2)(|x_j|^2 - |y_j|^2).
\end{align}
By \eqref{mixed2} and \eqref{mixed3}, the natural candidate for the norm $\| . \|_{H_n^0}$ is thus
\begin{align}
\label{mixed4}
N(\bDelta) = \sqrt{\sum\limits_{i<j} E^p_{ij} \left( |\Delta_{ij}|^2 - \Delta_{ii}\Delta_{jj} \right)}
\end{align}
for any $\bDelta \in H_n^0$. Later we will demonstrate that it is in fact the \emph{only} candidate, but first let us check whether $N$ is a well-defined Hilbertian norm in the first place.

To begin with, notice that, by Lemma \ref{mixedlem}, the radicand in \eqref{mixed4} is nonnegative for all $\bDelta \in H_n^0$ iff $E_{ij} = \|\bv_i-\bv_j\|^{2/p}$ for some set of vectors $\{\bv_1,\ldots,\bv_n\} \subset \sR^m$, which in turn allows us to rewrite $N$ as (cf. \eqref{mixedlem1})
\begin{align}
\label{mixed5}
N(\bDelta) = \sqrt{\sum_{i<j} \|\bv_i-\bv_j\|^2 |\Delta_{ij}|^2 + \left\| \sum\nolimits_k \Delta_{kk} \bv_k \right\|^2}.
\end{align}
If we now introduce the $\sR$-linear map $\Phi: H_n^0 \rightarrow H_n^0 \oplus \sR^m$ given by
\begin{align}
\label{Phi}
\Phi(\bDelta) := \left( \left(\tfrac{1}{\sqrt{2}} \|\bv_i-\bv_j\| \Delta_{ij}\right), \sum\nolimits_k \Delta_{kk} \bv_k \right),
\end{align}
and endow the space $H_n^0 \oplus \sR^m$ with the ``Frobenius $\oplus$ Euclidean'' inner product
\begin{align*}
\langle (\bA,\bw) | (\bB,\bu) \rangle_{\textup{F} \oplus \textup{E}} := \Tr \bA \bB + \bw \cdot \bu,
\end{align*}
then $N$ can be reexpressed simply as
\begin{align}
\label{mixed6}
N(\bDelta) = \|\Phi(\bDelta)\|_{\textup{F} \oplus \textup{E}}.
\end{align}
Thus, the necessary and sufficient condition for $N$ to be a Hilbertian norm is the injectivity of $\Phi$. But this, in turn, is equivalent to the affine independence of $\{\bv_1,\ldots,\bv_n\}$. Indeed, injectivity of $\Phi$ states that
\begin{align*}
\forall \bDelta \in H^0_n \quad \left[ \left( \|\bv_i-\bv_j\| \Delta_{ij}\right) = 0 \ \textnormal{and} \ \sum\nolimits_k \Delta_{kk} \bv_k = 0 \right] \ \Rightarrow \ \bDelta = 0,
\end{align*}
whereas the affine independence of $\{\bv_1,\ldots,\bv_n\}$ is characterized by
\begin{align*}
\forall \alpha_1,\ldots,\alpha_n \in \sR \quad \left[ \sum\nolimits_k \alpha_{k} = 0 \ \textnormal{and} \ \sum\nolimits_k \alpha_{k} \bv_k = 0 \right] \ \Rightarrow \ \alpha_1 = \ldots = \alpha_n = 0.
\end{align*}
To get the latter condition from the former one, simply take $\bDelta = \diag(\alpha_1,\ldots,\alpha_n)$ for any $\alpha_1,\ldots,\alpha_n$ satisfying the antecedent. Conversely, to obtain the former condition from the latter one, notice that, for any $\bDelta \in H^0_n$ satisfying the antecedent, the affine independence of $\{\bv_1,\ldots,\bv_n\}$ guarantees both that $\Delta_{ij} = 0$ for all $i \neq j$ (because all $\bv_i$'s are distinct) and that all $\Delta_{kk}$'s vanish as well. 

Let us now show that $N$ is the only Hilbertian norm on $H_n^0$ such that $N^2(\bx \bx^\dag - \by \by^\dag) = d_p^p(\bx,\by)$ for all $\bx,\by \in \PCn$. Indeed, suppose there exists a quadratic form $q$ on $H_n^0$ other than $N^2$ satisfying the above condition, and consider their difference $r := q - N^2$. The bilinear form associated with $r$ satisfies
\begin{align*}
b_r(\bx \bx^\dag - \bz \bz^\dag,\by \by^\dag - \bz \bz^\dag) = \tfrac{1}{2}\left( r(\bx \bx^\dag - \bz \bz^\dag) + r(\by \by^\dag - \bz \bz^\dag) - r(\bx \bx^\dag - \by \by^\dag) \right) = 0
\end{align*}
for any $\bx,\by,\bz \in \PCn$. But since for any fixed $\bz$ the matrices $\bx \bx^\dag - \bz \bz^\dag$ span the whole $H_n^0$ (by the spectral theorem), this implies that $b_r$ is the zero form and thus $r$ vanishes on the entire $H_n^0$, in contradiction with $q$ being distinct from $N^2$.

All in all, we have thus shown that $\|.\|_{H_n^0} = N$, and the distance matrix $(E_{ij})$ must be of the form $(\|\bv_i-\bv_j\|^{2/p})$ with $\{\bv_1,\ldots,\bv_n\}$ affinely independent for the whole construction to work and provide the isometric embedding of $\PCn$ into $\Omega_n$.
\end{proof}

Theorem \ref{mixedthm} justifies the following definition. For any $p \geq 2$ and any affinely independent $\{\bv_1,\ldots,\bv_n\} \subset \sR^m$, the map $\hat{d}_p$ given by
\begin{align}
\label{dhat}
\hat{d}_p(\rho,\sigma) & := \|\Phi(\rho - \sigma)\|^{2/p}_{\textup{F} \oplus \textup{E}} = \Big( \sum\limits_{i<j} \|\bv_i-\bv_j\|^2 \left( |\rho_{ij}-\sigma_{ij}|^2 - (\rho_{ii}-\sigma_{ii})(\rho_{jj}-\sigma_{jj}) \right) \Big)^{1/p}
\end{align}
for any $\rho,\sigma \in \Omega_n$ is a distance function that extends $d_p$, constructed upon the distance matrix $(E_{ij}) = (\|\bv_i-\bv_j\|^{2/p})$, to the mixed states.

In the remaining part of this section, let us study some basic features of $\hat{d}_p$.
\begin{remark}
The map $\hat{d}_p$ can also be expressed succinctly via (cf. \eqref{wasserstein} defining the quantum $p$-Wasserstein semidistance)
\begin{align}
\label{dhat2}
\hat{d}_p(\rho,\sigma) = \big({-\text{Tr}\left[\bE^p(\rho - \sigma) \wedge (\rho - \sigma)\right]}\big)^{1/p} \quad \textnormal{where} \quad \bE := \sum_{i < j} E_{ij} (\be_i \wedge \be_j)(\be_i \wedge \be_j)^\dagger.
\end{align}
Indeed, expanding the trace one has that
\begin{align*}
{-\text{Tr}\left[\bE^p(\rho - \sigma) \wedge (\rho - \sigma)\right]} = -\sum_{i < j} E_{ij}^p \langle \be_i \wedge \be_j | (\rho - \sigma)\be_i \wedge (\rho - \sigma)\be_j \rangle
\end{align*}
and a straightforward calculation shows that $\langle \be_i \wedge \be_j | (\rho - \sigma)\be_i \wedge (\rho - \sigma)\be_j \rangle = (\rho_{ii}-\sigma_{ii})(\rho_{jj}-\sigma_{jj})-|\rho_{ij}-\sigma_{ij}|^2$, which together with $E_{ij}^p = \|\bv_i-\bv_j\|^2$ yields \eqref{dhat}.
\end{remark}

\begin{proposition}
For any $p \geq 2$ the map $\hat{d}_p$ is jointly $2/p$-convex, i.e., it satisfies
\begin{align}
\label{quasiconvex}
\hat{d}_p(t\sigma_1 + (1-t)\sigma_2,t\rho_1 + (1-t)\rho_2) \leq t^{2/p}\hat{d}_p(\sigma_1,\rho_1) + (1-t)^{2/p}\hat{d}_p(\sigma_2,\rho_2)
\end{align}
for any $\sigma_1,\sigma_2,\rho_1,\rho_2 \in \Omega_n$ and $t \in [0,1]$. In particular, $\hat{d}_2$ is jointly convex.
\end{proposition}
\begin{proof}
Expressing both sides of \eqref{quasiconvex} with the help of \eqref{dhat}, one can write that
\begin{align*}
\textnormal{LHS} & = \|t\Phi(\sigma_1-\rho_1) - (t-1)\Phi(\sigma_2-\rho_2)\|^{2/p}_{\textup{F} \oplus \textup{E}} \leq \|t\Phi(\sigma_1-\rho_1)\|^{2/p}_{\textup{F} \oplus \textup{E}} + \|(1-t)\Phi(\sigma_2-\rho_2)\|^{2/p}_{\textup{F} \oplus \textup{E}}
\\
& = t^{2/p}\|\Phi(\sigma_1-\rho_1)\|^{2/p}_{\textup{F} \oplus \textup{E}} + (1-t)^{2/p}\|\Phi(\sigma_2-\rho_2)\|^{2/p}_{\textup{F} \oplus \textup{E}} = \textnormal{RHS},
\end{align*}
where we have used the triangle inequality for the distance function on $H_n^0 \oplus \sR^m$ given by $((\bA,\bw),(\bB,\bu)) \mapsto \|(\bA,\bw)-(\bB,\bu)\|^{2/p}_{\textup{F} \oplus \textup{E}}$.
\end{proof}

\begin{corollary}
For any $p \geq 2$ the map $\hat{d}_p$ is jointly \emph{quasi}convex, i.e.,
\begin{align}
\label{quasiconvex1}
\hat{d}_p(t\sigma_1 + (1-t)\sigma_2,t\rho_1 + (1-t)\rho_2) \leq \max\left\{ \hat{d}_p(\sigma_1,\rho_1), \hat{d}_p(\sigma_2,\rho_2) \right\}
\end{align}
for any $\sigma_1,\sigma_2,\rho_1,\rho_2 \in \Omega_n$ and $t \in [0,1]$.
\end{corollary}

Concerning the symmetries of $\hat{d}_p$, we generalize Proposition \ref{prop_sym}.
\begin{proposition}
\label{mixed_sym}
For any $\bU \in \textup{U}(n)$ the following conditions are equivalent
\begin{enumerate}[(i)]
\item $\hat{d}_p(\bU\rho\bU^\dag,\bU\sigma\bU^\dag) = \hat{d}_p(\rho,\sigma)$ for any $\rho,\sigma \in \Omega_n$.
\item $\bE$ commutes with $\bU \wedge \bU$.
\end{enumerate}
\end{proposition}
\begin{proof}
The implication \emph{(i)} $\Rightarrow$ \emph{(ii)} follows from the corresponding implication in Proposition \ref{prop_sym}. To show the converse implication, one can employ \eqref{dhat2} and write that, for any $\rho,\sigma \in \Omega_n$,
\begin{align*}
\hat{d}^p_p(\bU\rho\bU^\dag,\bU\sigma\bU^\dag) & = {-\text{Tr}\left[\bE^p(\bU \wedge \bU)((\rho - \sigma) \wedge (\rho - \sigma))(\bU^\dag \wedge \bU^\dag)\right]}
\\
& = {-\text{Tr}\left[\bE^p(\rho - \sigma) \wedge (\rho - \sigma)\right]} = \hat{d}^p_p(\rho,\sigma),
\end{align*}
employing the cyclic property of the trace and the assumed commutation of $\bE$ and $\bU \wedge \bU$.
\end{proof}

The distances $\hat{d}_p$ can be regarded as generalizing the Frobenius distance raised to the power $2/p$. Indeed, the distance matrix $(E_{ij}) = (1-\delta_{ij})$ (whose entries are indeed of the required form $\|\bv_i - \bv_j\|^{2/p}$, as realized by taking, e.g., $\bv_i := \be_i/\sqrt{2}$) yields exactly
\begin{align*}
\hat{d}_p(\sigma,\rho) & = \Big( \tfrac{1}{2} \sum_{i,j} \left(|\sigma_{ij}-\rho_{ij}|^2-(\sigma_{ii}-\rho_{ii})(\sigma_{jj}-\rho_{jj})\right) \Big)^{1/p}
\\
& = \Big( \tfrac{1}{2} \sum_{i,j} |\sigma_{ij}-\rho_{ij}|^2 \Big)^{1/p} = 2^{-1/p}\|\sigma-\rho\|^{2/p}_\textup{F}.
\end{align*}
In fact, the distances $\hat{d}_p$ are uniformly equivalent to such a distance, as attested by the theorem below (cf. Remark \ref{rem_poltora}). Before we state it, notice that for any distance matrix of the form $(\|\bv_i - \bv_j\|^{2/p})$, the vectors $\bv_1,\ldots,\bv_n$ can be assumed to satisfy $\sum_i \bv_i = 0$ without loss of generality.
\begin{theorem}
The map $\hat{d}_p$ given by \eqref{dhat} satisfies, for some positive constants $C,c$,
\begin{align}
\label{equiv}
c\|\sigma-\rho\|^{2/p}_\textup{F} \leq \hat{d}_p(\sigma,\rho) \leq C\|\sigma-\rho\|^{2/p}_\textup{F}
\end{align}
for any $\rho,\sigma \in \Omega_n$. Furthermore, let $\bv_1,\ldots,\bv_n$ be the vectors defining $\hat{d}_p$ such that $\sum_i \bv_i = 0$, and let $\bG$ be their Gram matrix with eigenvalues $\lambda_1, \ldots, \lambda_n$. Then the optimal $C$ in \eqref{equiv} is the largest number among the numbers $2^{-1/p} \|\bv_i - \bv_j\|^{2/p}$ and $\lambda_k^{1/p}$, whereas the optimal $c$ is the smallest positive number among them.
\end{theorem}
\begin{proof}
By \eqref{dhat}, the optimal constants $C,c$ are given by
\begin{align}
\label{equiv1}
C = \left( \max_{\rho \neq \sigma} \frac{\|\Phi(\rho - \sigma)\|_{\textup{F} \oplus \textup{E}}}{\|\rho - \sigma\|_\textup{F}} \right)^{2/p} \quad \textnormal{and} \quad c = \left( \min_{\rho \neq \sigma} \frac{\|\Phi(\rho - \sigma)\|_{\textup{F} \oplus \textup{E}}}{\|\rho - \sigma\|_\textup{F}} \right)^{2/p}.
\end{align}
We claim that the set $\cS := \{\rho - \sigma \, | \, \rho,\sigma \in \Omega_n, \rho \neq \sigma\}$, over which we are optimizing, can be characterized as
\begin{align*}
\cS = \{ \bDelta \in H^0_n \, | \, 0 < \|\bDelta\|_1 \leq 2 \},
\end{align*}
where $\|.\|_1$ denotes the trace norm. 

Indeed, for any two distinct density matrices $\rho,\sigma$, their difference is traceless and satisfies $0 < \|\rho - \sigma\|_1 \leq \|\rho\|_1 + \|\sigma\|_1 = 2$. Conversely, let $\bDelta \in H^0_n$ satisfy $0 < \|\bDelta\|_1 \leq 2$. Write $\bDelta = \bU^\dag(\bD_+-\bD_-)\bU$, where $\bU$ is unitary and $\bD_\pm$ are the positive and negative parts of $\bDelta$'s diagonal form. Their common trace is $\tau = \|\bDelta\|_1/2 \leq 1$. But then $\rho := \bU^\dag \bD_+ \bU + \tfrac{1-\tau}{n}\bI$ and $\sigma := \bU^\dag \bD_- \bU + \tfrac{1-\tau}{n}\bI$ are well-defined, distinct mixed states such that $\bDelta = \rho - \sigma$.

With thus proven claim in mind, (\ref{equiv1}) becomes
\begin{align}
\label{equiv2}
C = \left( \max_{\bDelta \in \cS} \frac{\|\Phi(\bDelta)\|_{\textup{F} \oplus \textup{E}}}{\|\bDelta\|_\textup{F}} \right)^{2/p} \quad \textnormal{and} \quad c = \left( \min_{\bDelta \in \cS} \frac{\|\Phi(\bDelta)\|_{\textup{F} \oplus \textup{E}}}{\|\bDelta\|_\textup{F}} \right)^{2/p}.
\end{align}
Notice that one can, in fact, ignore the condition $\|\bDelta\|_1 \leq 2$, because in the above fractions $\bDelta$ can always be renormalized, and we can simply write
\begin{align}
\label{equiv3}
C = \bigg( \max_{\substack{\bDelta \in H_n^0 \\ \|\bDelta\|_\textup{F} = 1}} \|\Phi(\bDelta)\|_{\textup{F} \oplus \textup{E}} \bigg)^{2/p} \quad \textnormal{and} \quad c = \bigg( \min_{\substack{\bDelta \in H_n^0 \\ \|\bDelta\|_\textup{F} = 1}} \|\Phi(\bDelta)\|_{\textup{F} \oplus \textup{E}} \bigg)^{2/p}.
\end{align}
In order to calculate both above optima, it is crucial to observe that $\Phi$ splits with respect to the orthogonal decomposition $H_n^0 = H_n^\textup{h} \oplus H_n^{0,\textup{d}}$, where $H_n^\textup{h}$ denotes the subspace of hollow Hermitian matrices and $H_n^{0,\textup{d}}$ is the subspace of real traceless diagonal matrices. Concretely,
\begin{align}
\Phi = \Phi_\textup{h} \oplus \Phi_\textup{d},
\end{align}
where $\Phi_\textup{h}: H_n^\textup{h} \rightarrow H^0_n$ is defined via $\Phi_\textup{h}(\bA) := (\tfrac{1}{\sqrt{2}} \|\bv_i-\bv_j\| A_{ij})$ and $\Phi_\textup{d}: H_n^{0,\textup{d}} \rightarrow \sR^m$ is given by $\Phi_\textup{d}(\bD) := \sum\nolimits_k D_{kk} \bv_k$. This splitting allows us to perform the optimizations separately on each of the subspaces $H_n^\textup{h}$ and $H_n^\textup{0,d}$, with (\ref{equiv3}) becoming
\begin{align}
\label{equiv4}
& C = \Big( \max\big\{ \max_{\substack{\bA \in H_n^\textup{h} \\ \|\bA\|_\textup{F} = 1}} \|\Phi_\textup{h}(\bA)\|_\textup{F}, \max_{\substack{\bD \in H_n^{0,\textup{d}} \\ \|\bD\|_\textup{F} = 1}} \|\Phi_\textup{d}(\bD)\| \big\} \Big)^{2/p}
\\
\nonumber
\textnormal{and} \quad & c = \Big( \min\big\{ \min_{\substack{\bA \in H_n^\textup{h} \\ \|\bA\|_\textup{F} = 1}} \|\Phi_\textup{h}(\bA)\|_\textup{F}, \min_{\substack{\bD \in H_n^{0,\textup{d}} \\ \|\bD\|_\textup{F} = 1}} \|\Phi_\textup{d}(\bD)\| \big\} \Big)^{2/p}.
\end{align}
It is not hard to see that
\begin{align*}
\max_{\substack{\bA \in H_n^\textup{h} \\ \|\bA\|_\textup{F} = 1}} \|\Phi_\textup{h}(\bA)\|_\textup{F} = \tfrac{1}{\sqrt{2}} \max_{i,j} \|\bv_i-\bv_j\| \quad \textnormal{and} \quad \min_{\substack{\bA \in H_n^{0,\textup{h}} \\ \|\bA\|_\textup{F} = 1}} \|\Phi_\textup{h}(\bA)\|_\textup{F} = \tfrac{1}{\sqrt{2}} \min_{i \neq j} \|\bv_i-\bv_j\|.
\end{align*}

Concerning the other maximum, it is worthwhile to reinterpret the diagonal traceless matrix $\bD$ as a vector $\bba$ orthogonal to $\bjed$ and observe that
\begin{align*}
\max_{\substack{\bD \in H_n^{0,\textup{d}} \\ \|\bD\|_\textup{F} = 1}} \|\Phi_\textup{d}(\bD)\| = \max_{\substack{\bba \bot \bjed \\ \|\bba\|  = 1}} \big\|\sum\nolimits_k a_k \bv_k\big\| = \max_{\substack{\bba \bot \bjed \\ \|\bba\| = 1}} \sqrt{\sum\nolimits_{k,l} a_k a_l \bv_k \cdot \bv_l} = \max_{\substack{\bba \bot \bjed \\ \|\bba\| = 1}} \sqrt{\bba^\top \bG \bba} = \max_{\substack{\bba \, \bot \ker\bG \\ \|\bba\| = 1}} \sqrt{\bba^\top \bG \bba},
\end{align*}
where in the last equality we have used the fact that $\bjed$ spans the null space of $\bG$ (what follows from $\sum_i \bv_i = 0$). The rightmost maximum is nothing but the square root of the largest eigenvalue of $\bG$.

For the remaining minimum we obtain, completely analogously,
\begin{align*}
\min_{\substack{\bD \in H_n^{0,\textup{d}} \\ \|\bD\|_\textup{F} = 1}} \|\Phi_\textup{d}(\bD)\| = \min_{\substack{\bba \, \bot \ker\bG \\ \|\bba\| = 1}} \sqrt{\bba^T \bG \bba},
\end{align*}
which is equal to the square root of the smallest \emph{positive} eigenvalue of $\bG$, thus concluding the proof.
\end{proof}

\begin{remark}
Using the elementary inequalities between the Schatten norms, we can now also compare $\hat{d}_p$ with the trace distance $T(\rho,\sigma) := \tfrac{1}{2}\|\rho - \sigma\|_1$, namely
\begin{align*}
\left( \tfrac{2}{\sqrt{n}} \right)^{2/p} c T^{2/p}(\rho,\sigma) \leq \hat{d}_p(\rho,\sigma) \leq 2^{2/p} C T^{2/p}(\rho,\sigma).
\end{align*}
Furthermore, the Fuchs--van de Graaf inequalities \cite{Fuchs} can now be used to compare $\hat{d}_p$ with other standard distances on $\Omega_n$ such as the Bures distance $B$, obtaining
\begin{align*}
c n^{-1/p} B^{4/p}(\rho,\sigma) \leq \hat{d}_p(\rho,\sigma) \leq 2^{2/p} C B^{2/p}(\rho,\sigma).
\end{align*}
\end{remark}

\begin{remark}
In the context of quantum information theory, it is natural to ask how the distances $\hat{d}_p$ behave under completely positive trace-preserving (CPTP) maps, which model quantum channels. In particular, one may ask whether such maps are monotone (nonexpansive) with respect to $\hat{d}_p$. The answer is negative. This is unsurprising, since $\hat{d}_p$ generalizes the Frobenius distance, and CPTP maps need not be monotone with respect to the latter. Moreover, in contrast to the Frobenius distance, even unitary channels need not be monotone with respect to $\hat{d}_p$, as shown by the following counterexample.

Let $n=3$ and take $\bv_1 := (0,0)$, $\bv_2 := (1,0)$ and $\bv_3 := (0,2)$. Also let $\bx := \tfrac{1}{\sqrt{2}}(\be_1 + \be_2)$ and $\by := \tfrac{1}{\sqrt{2}}(\be_1 - \be_2)$ and consider the CPTP map $\rho \mapsto \bP\rho\bP^\dag$, where $\bP \in \textup{U}(3)$ is the permutation matrix corresponding to the transposition $2 \leftrightarrow 3$. For any $p \geq 2$ the distance $\hat{d}_p$ constructed upon $\{\bv_1,\bv_2,\bv_3\}$ via \eqref{dhat} satisfies 
\begin{align*}
& \hat{d}_p(\bx\bx^\dag, \by\by^\dag) = d_p(\bx,\by) = \|\bv_1 - \bv_2\|^{2/p} = 1,
\\
\textnormal{and} \quad & \hat{d}_p(\bP\bx\bx^\dag\bP^\dag, \bP\by\by^\dag\bP^\dag) = d_p(\bP\bx,\bP\by) = \|\bv_1 - \bv_3\|^{2/p} = 4^{1/p},
\end{align*}
thus violating the monotonicity condition.
\end{remark}

\section*{Acknowledgments}
We are grateful to Shmuel Friedland, Karol \.{Z}yczkowski and Micha\l{} Eckstein for their insightful comments and helpful discussions, as well as to the anonymous referees, whose remarks and suggestions greatly improved the manuscript. This work was supported by the National Science Center, Poland, under the contract number 2023/50/E/ST2/00472 and under the contract number 2021/03/Y/ST2/00193, within the QuantERA II Programme under Grant Agreement No 101017733, and by the European Union under ERC Advanced Grant \emph{TAtypic}, project number 101142236.

\bibliographystyle{habbrv}
\bibliography{trianglerefs1}

\begin{thebibliography}{10}

\bibitem{Beatty}
E.~Beatty.
\newblock {Wasserstein} {D}istances on {Q}uantum {S}tructures: an {O}verview,
  2025, 2506.09794.

\bibitem{BeattyFranca}
E.~Beatty and D.~Stilck~Fran{\c{c}}a.
\newblock Order {$p$} quantum {W}asserstein distances from couplings.
\newblock {\em Ann. Henri Poincar{\'e}}, 27(3):787--845, 2026.

\bibitem{BZ17}
I.~Bengtsson and K.~Życzkowski.
\newblock {\em Geometry of Quantum States: An Introduction to Quantum
  Entanglement}.
\newblock Cambridge University Press, 2 edition, 2017.

\bibitem{QML}
J.~Biamonte, P.~Wittek, N.~Pancotti, P.~Rebentrost, N.~Wiebe, and S.~Lloyd.
\newblock Quantum machine learning.
\newblock {\em Nature}, 549:195--202, Sept. 2017.

\bibitem{BEZ22}
R.~Bistro\'{n}, M.~Eckstein, and K.~\.{Z}yczkowski.
\newblock Monotonicity of a quantum 2-{W}asserstein distance.
\newblock {\em J. Phys. A}, 56, 2023.

\bibitem{Energy_distance}
R.~Bistro\'{n}, M.~Eckstein, and K.~\.{Z}yczkowski.
\newblock {Hamiltonian-induced distance in the space of quantum states}.
\newblock In preparation, 2026.

\bibitem{llaproc}
R.~Bistroń, M.~Eckstein, S.~Friedland, T.~Miller, and K.~Życzkowski.
\newblock A new class of distances on complex projective spaces.
\newblock {\em Linear Algebra Appl.}, 721:577--611, 2025.

\bibitem{Borsoni}
T.~Borsoni.
\newblock Metric convex extensions and optimal transport on classical and
  quantum structures, 2025, 2512.01722.

\bibitem{CGP20}
E.~Caglioti, F.~Golse, and T.~Paul.
\newblock Quantum optimal transport is cheaper.
\newblock {\em J. Stat. Phys.}, 181:149--162, 2020.

\bibitem{CarlenMaas}
E.~A. Carlen and J.~Maas.
\newblock An analog of the {$2$}-{Wasserstein} metric in non-commutative
  probability under which the fermionic {Fokker--Planck} equation is gradient
  flow for the entropy.
\newblock {\em Commun. Math. Phys.}, 331(3):887--926, 2014.

\bibitem{CEFZ21}
S.~Cole, M.~Eckstein, S.~Friedland, and K.~\.{Z}yczkowski.
\newblock On quantum optimal transport.
\newblock {\em Math. Phys. Anal. Geom.}, 26, 2023.

\bibitem{LHK13}
J.~A. {De Loera}, R.~Hemmecke, and M.~K\"{o}ppe.
\newblock {\em Algebraic and Geometric Ideas in the Theory of Discrete
  Optimization}.
\newblock SIAM, 2013.

\bibitem{PMTL21}
G.~{De Palma}, M.~Marvian, D.~Trevisan, and S.~Lloyd.
\newblock The quantum {W}asserstein distance of order $1$.
\newblock {\em IEEE Trans. Inf. Theory}, 67:6627--6643, 2021.

\bibitem{PT21}
G.~{De Palma} and D.~Trevisan.
\newblock Quantum optimal transport with quantum channels.
\newblock {\em Ann. Henri Poincar\'e}, 22:3199--3234, 2021.

\bibitem{QSense}
C.~L. Degen, F.~Reinhard, and P.~Cappellaro.
\newblock Quantum sensing.
\newblock {\em Rev. Mod. Phys.}, 89, 2017.

\bibitem{Duv22}
R.~Duvenhage.
\newblock Quadratic {W}asserstein metrics for von {N}eumann algebras via
  transport plans.
\newblock {\em J. Oper. Theory}, 88:289--308, 2022.

\bibitem{FECZ21}
S.~Friedland, M.~Eckstein, S.~Cole, and K.~\.{Z}yczkowski.
\newblock Quantum {M}onge--{K}antorovich problem and transport distance between
  density matrices.
\newblock {\em Phys. Rev. Lett.}, 129, 2022.

\bibitem{Fuchs}
C.~A. Fuchs and J.~van~de Graaf.
\newblock Cryptographic distinguishability measures for quantum-mechanical
  states.
\newblock {\em IEEE Trans. Inf. Theory}, 45(4):1216--1227, 1999.

\bibitem{QMetro}
V.~Giovannetti, S.~Lloyd, and L.~Maccone.
\newblock Advances in quantum metrology.
\newblock {\em Nat. Photonics}, 5:222--229, 2011.

\bibitem{GP18}
F.~Golse and T.~Paul.
\newblock Wave packets and the quadratic {M}onge--{K}antorovich distance in
  quantum mechanics.
\newblock {\em C. R. Math.}, 356:177--197, 2018.

\bibitem{Kan48}
L.~V. Kantorovich.
\newblock On a problem of {M}onge.
\newblock {\em J. Math. Sci.}, 133:1383, 2006.

\bibitem{Keyl}
M.~Keyl.
\newblock Fundamentals of quantum information theory.
\newblock {\em Phys. Rep.}, 369:431--548, 2002.

\bibitem{KdPMLL22}
B.~T. Kiani, G.~{De Palma}, M.~Marvian, Z.-W. Liu, and S.~Lloyd.
\newblock Learning quantum data with the quantum {E}arth {M}over’s distance.
\newblock {\em Quantum Sci. Technol.}, 7, 2022.

\bibitem{MillerComment}
T.~Miller.
\newblock Comment on '{Q}uantum {M}onge--{K}antorovich problem and transport
  distance between density matrices', 2026, 2607.07764.

\bibitem{MillerSep}
T.~Miller.
\newblock Metricity of separable quantum optimal transport with the
  {H}ilbert--{S}chmidt cost, 2026, 2609.03769.

\bibitem{NielsenChuang}
M.~A. Nielsen and I.~L. Chuang.
\newblock {\em Quantum Computation and Quantum Information}.
\newblock Cambridge University Press, 2010.

\bibitem{Schneider}
R.~Schneider.
\newblock {\em Convex Bodies: The Brunn--Minkowski Theory}.
\newblock Cambridge University Press, 2 edition, 2014.

\bibitem{Schoenberg}
I.~J. Schoenberg.
\newblock Metric spaces and positive definite functions.
\newblock {\em Trans. Am. Math. Soc.}, 44(3):522--536, 1938.

\bibitem{TothPitrik23}
G.~T{\'o}th and J.~Pitrik.
\newblock Quantum {W}asserstein distance based on an optimization over
  separable states.
\newblock {\em Quantum}, 7:1143, 2023.

\bibitem{TothPitrik26}
G.~T{\'o}th and J.~Pitrik.
\newblock Quantum {W}asserstein distance and its relation to several types of
  fidelities, 2025, 2506.14523.

\bibitem{Invitation}
D.~Trevisan.
\newblock Quantum optimal transport: an invitation.
\newblock {\em Bollettino dell'Unione Matematica Italiana}, 18(1):347--360,
  Mar. 2025.

\bibitem{U_learn}
A.~Usui, G.~Abad-L\'{o}pez, H.~k. SV, A.~Sanpera, and S.~S. Bhattacharya.
\newblock Entanglement-assisted {H}amiltonian dynamics learning, 2026.

\bibitem{Wells}
R.~O. Wells.
\newblock {\em Differential Analysis on Complex Manifolds}.
\newblock Springer, 3 edition, 2008.

\bibitem{ZYYY22}
L.~Zhou, N.~Yu, S.~Ying, and M.~Ying.
\newblock Quantum earth mover's distance, no-go quantum
  {K}antorovich--{R}ubinstein theorem, and quantum marginal problem.
\newblock {\em J. Math. Phys.}, 63, 2022.

\bibitem{Hamming_distance}
M.~Ziobro, F.~Ungeheuer, R.~Bistro\'{n}, and K.~\.{Z}yczkowski.
\newblock {Quantum Wasserstein GAN enhanced by Hamming geometry}.
\newblock In preparation, 2026.

\bibitem{ZS01}
K.~\.{Z}yczkowski and W.~S{\l}omczy{\'n}ski.
\newblock The {M}onge metric on the sphere and geometry of quantum states.
\newblock {\em J. Phys. A}, 34, 2001.

\end{thebibliography}
\end{document}